\documentclass[journal]{IEEEtran}

\usepackage[T1]{fontenc} 
\usepackage[utf8]{inputenc}
\usepackage{pgfplots}
\usepackage{tikz}
\usepackage{pgfplots}
\pgfplotsset{compat=1.8}
\usepackage{pgfplotstable}
\usepgfplotslibrary{groupplots}
\usepackage{afterpage}
\usepackage{siunitx}
\ifCLASSINFOpdf
\else
\fi
\usepackage{url}

\usepackage{amsfonts}
\usepackage{amsmath}
\usepackage{subfig}
\usepackage{todonotes}

\usepackage{tikz}

\usetikzlibrary{arrows,automata}
\usetikzlibrary{positioning}

\tikzset{
	state/.style={
		rectangle,
		rounded corners,
		draw=black, very thick,
		minimum height=2em,
		inner sep=2pt,
		text centered,
	},
}

\newtheorem{re}{Remark}
\newtheorem{as}{Assumption}

\begin{document}

\title{Generalized Score Distribution}

\author{Lucjan~Janowski, Bogdan~Ćmiel, Krzysztof~Rusek, Jakub~Nawała, Zhi~Li
	\thanks{B.~Ćmiel is with the Department of Applied Mathematics, AGH University of Science and Technology, Poland.}
	\thanks{L.~Janowski, J.~Nawała, and K.~Rusek are with the Department of Telecommunictions of AGH University of Science and Technology, Poland. e-mail: janowski@kt.agh.edu.pl}
	\thanks{Z.~Li is with Netflix}}

\maketitle

\begin{abstract}
A class of discrete probability distributions contains distributions with limited support, i.e. possible argument values are limited to a set of numbers (typically consecutive). Examples of such data are results from subjective experiments utilizing the Absolute Category Rating (ACR) technique, where possible answers (argument values) are $\{1, 2, \cdots, 5\}$ or typical Likert scale $\{-3, -2, \cdots, 3\}$. An interesting subclass of those distributions are distributions limited to two parameters: describing the mean value and the spread of the answers, and having no more than one change in the probability monotonicity. In this paper we propose a general distribution passing those limitations called Generalized Score Distribution (GSD). The proposed GSD covers all spreads of the answers, from very small, given by the Bernoulli distribution, to the maximum given by a Beta Binomial distribution. We also show that GSD correctly describes subjective experiments scores from video quality evaluations with probability of 99.7\%. A Google Collaboratory website with implementation of the GSD estimation, simulation, and visualization is provided.
\end{abstract}


\section{Introduction}

\IEEEPARstart{S}{ubjective} experiments let us collect users' opinions about a specific system. Such experiment can be seen as a measuring device. Any measuring device provides certain precision. In this article we analyze subjects'/testers'/users' answers, in order to better understand the random process generating those answers. The obtained result is general and covers interesting class of discrete distributions with limited support, and two parameters describing the mean and the answer's spread. 

Our main goal is to check what is the distribution of answers provided by subjects. We limit our considerations to popular subjective experiments using discrete five point scale $\{1, 2, \cdots, 5\}$. For this scale, numerous different databases are available therefore we compare the theoretical and analytical results. Nevertheless, the proposed model is general and in the Appendix~\ref{sec:ap:n} we describe the generalization to $M$ point scale.

Knowing the answers' distribution provides important information about the answering process. Further modeling of this process can lead to better understanding of experiments involving subjects \cite{Tobias_no_silver_bullet_2017}. The ultimate goal is to make data analysis of those experiments more precise. Better understanding which factors influence the obtained answer allows to remove, or compensate, those aspects in the data analysis. 

The main contribution of this paper is to propose a new Generalized Score Distribution (GSD), and provide a strong evidence that GSD describes the answering process in a video quality test correctly. We strongly believe that GSD distribution can be useful in different fields of science, since it is a generalization of the existing and widely used distributions. For the proposed distribution the estimation algorithm\footnote{The implementation can be found here \url{https://colab.research.google.com/drive/1ioM4JqUaEA8fHJH9V-0iphHt4C9zmI0i}} is provided and analyzed by extensive simulation study. The analysis of different existing video quality databases allow us to estimate typical parameters of this model. Those results can be used as a prior distribution of the GSD parameters in the future analysis such as Bayes estimator.

In order to validate GSD in the context of subjective experiments we have used four different databases, with more than 1,800 PVSs (Processed Video Sequences, i.e. scored sequences). The analyzed PVSs have different number of votes from typical 24 up to 213. The final score shows that GSD describes the answers correctly with probability of 99.7\%.  

In the next section we describe the related work. Section~\ref{sec:model} describes the proposed distribution with the estimation process detailed in Section~\ref{sec:estimation}. Section~\ref{sec:datasets} provides short descriptions of the data sets used. Section~\ref{sec:analysis} presents the results. The last section concludes the paper.


\section{Related Work}
\label{sec:relatedwork}

Subject answers analysis is a broad topic considered in numerous publications starting from ITU standards to technical publications provided by companies. The standard focusing on the objective metrics evaluation is ITU-T P.1401 \cite{ITUP1401}. Its main goal is to describe correct way of analyzing subjective data. Similarly, a recent publication \cite{Brunnstrom2018} describes the problem of correct way of comparing groups within subjective experiment. 

Beyond correctness in the literature we can find interesting proposals for other than typical, focused on MOS, analysis. In \cite{Hossfeld2016} a number of different metrics related to user behavior and service acceptance are described. In \cite{Janowski2009} a method to find each answer probability is suggested. More recently similar approach to model each answer probability, instead of simple mean (MOS), was proposed in \cite{Seufert_Fundamental_2019}. Other publications are focused on modeling Quality of Experience in general e.g. \cite{Fiedler2010}. The model we propose in this paper is in line with those analysis as the model is discrete. 

Not many publications focus on the answering process, especially the precision of the obtained score. An interesting analysis is shown in \cite{Hossfeld2011}, where the relation between MOS and the standard deviation of scores is studied. In the paper a unique parameter HSE\footnote{In the original paper the parameter was named SOS but we think that HSE (from Hossfeld, Schatz, and Egger) will be a better, unambiguous name. SOS is traditionally used to describe standard deviation of scores.}, describing the test difficulty was proposed. Another interesting proposal analyzing the subjective experiment precision is \cite{Hossfeld2018} where different confidence intervals are studied. 

The analysis of the subjective experiment leads to proposing two subject models in \cite{Janowski2015} and \cite{Li2017}. Those models are discussed in more detail further in the paper. User model allows to extend analysis of the subjective experiments like the one described in \cite{Tobias_no_silver_bullet_2017, Kumcu_2017_Performace_of_four}, or make the existing analysis more precise \cite{Improve_analysis_2018, Freitas_2018_Performance}. The user model is modified in \cite{User_Model_for_JND_2018, Tasaka_2017_Bayesian_Hierarchical, Pablo_AMP_2019} which leads to new results, e.g. showing that the content has significant influence on both variability and mean of the obtained answers \cite{Pablo_AMP_2019}. 

The new subject model proposed in this paper extends the existing analysis by introducing discrete distribution. The previous publications used continuous process with discretization and censoring (clipping). As the next section describes, this introduces certain error. Therefore, the proposed discrete model, not having this error, is a better solution. 

In this paper we do not compare our model to models used in different fields working on quality. It is left as an interesting future research. Comparing to those modes is difficult, since all those models need a different data, which do not exists commonly in the video quality community. An example could be food industry using ``Panel Check'' \cite{Romano_Correcting_2008, Tomic_Visualization_2007}, the method proposed by Pane Check relay strongly on the lack of tied answers. For the five point scale used in video quality evaluation, small number of tied answers is a strong assumption.

In psychology the signal detection theory (SDT) is used in the context of subjective quality scores analysis \cite{Maniscalco_SDT_2012, Fleming_Self-evaluation_2017}. Such analysis can be complicated taking into account different influencing factors \cite{King_A_model_of_subjective_2014}. This is especially an interesting connection between QoE and psychology, nevertheless, again a slightly different data are needed. To use SDT we need a recognition and a quality score, in the video quality analysis case only quality scores are available. 

It is also worth mentioning that the distribution proposed in this paper can be used for modeling an answering in Likert scale surveys. For example, in \cite{NUMBER_OF_RESPONSE} the authors consider the impact of the number of response categories on the reliability of the measurements. In the model of answers provided in equation (1) of \cite{NUMBER_OF_RESPONSE} one can consider GSD distribution of the random error to estimate the latent unobserved true score. The same approach can also be used for example in \cite{PSYCHOMETRIC_TESTING} where the problem of measuring patients experience in hospitals is considered. 

\section{Subject Answers as a Random Variable}
\label{sec:model}

Assuming that a test is conducted using five point discrete scale the subject answer $U$ has a multinomial distribution given by:
\begin{equation}
P(U = s) = p_{s}, \textrm{ where } \sum_{s=1}^{5}p_s = 1
\end{equation}

Such description of a subject answers distribution is general but has four different parameters, coming from five probabilities. 

In general we can describe the subject answer as a function: 
\begin{equation}
U = \psi + \epsilon,
\end{equation}
where $\psi$ is the true quality and $\epsilon$ is an error term. An algorithm predicting quality should aim in estimating $\psi$. Still, the error distribution is important and should be modeled. The error term represents precision of $\psi$ estimation. Note that the variance term (represented by $\epsilon$) cannot be too complicated. In total, we have four different probabilities. Therefore, we would like to use a model for which the variance is described by a single parameter. We can then use it to predict the true quality. In turn, accuracy of its estimation is denoted by the variance parameter.

For simplicity, we only consider a single sequence (i.e., PVS) analysis. In general, we consider a model, where an answer for a  sequence is a random variable $U$ drawn from a distribution:
\begin{equation}
U \sim F(\psi, \theta),
\end{equation}
where $\psi$ is the true quality, $\theta$ is a parameter describing answers spread, and $F()$ a cumulative distribution function. However, in the text we mark $\theta$ differently, depending on the model we are considering. This notation makes the description clearer. 

Our notation convention generally follows guidelines of SAM (Statistical Analysis Methods) of VQEG (Video Quality Expert Group) group, described in \cite{Janowski2019NotationFS}. Nevertheless, we introduce extensions to selected symbols.

\subsection{Continuous Model}

The models proposed by \cite{Janowski2015} and \cite{Li2017} describe a subject answer as a continuous normal distribution with certain mean $\psi$, which is true quality, and standard deviation $\sigma$ describing the error. Therefore, the subject answer is $O \sim \mathcal{N}(\psi, \sigma)$. Since a subjective scale is discrete we cannot observe $O$. To convert a continuous answer to a discrete one, a subject makes discretization and censoring (clipping). This process converts continuous random variable $O$ to discrete variable $U$. We can calculate each answer probability (i.e., $U$ distribution), as a function of $\psi$ and $\sigma$ by the below equations:
\begin{equation}
P(U = s) = \int_{s-0.5}^{s+0.5}\frac{1}{2\pi \sigma}e^{-\frac{(x-\psi)^2}{2\sigma^2}}
\end{equation}
for $s = \{2,3,4\}$ and 
\begin{equation}
P(U = 1) = \int_{-\infty}^{1.5}\frac{1}{2\pi \sigma}e^{-\frac{(x-\psi)^2}{2\sigma^2}}
\end{equation}
\begin{equation}
P(U = 5) = \int_{4.5}^{\infty}\frac{1}{2\pi \sigma}e^{-\frac{(x-\psi)^2}{2\sigma^2}}
\end{equation}

\begin{figure}
	\centering
	\input{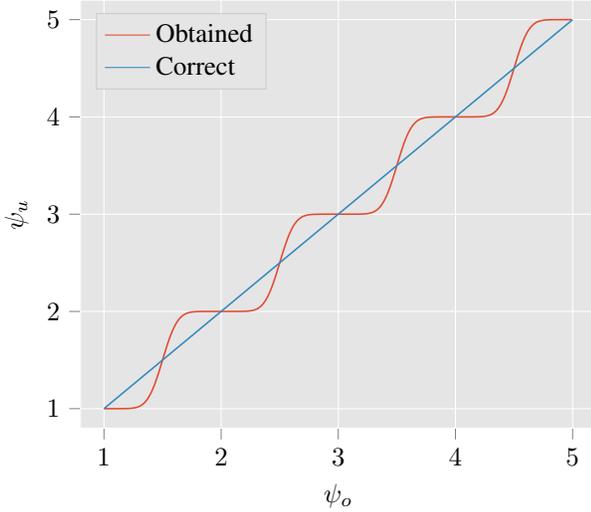}
	\caption{The difference between $\psi_o$ (the true quality for continuous model) and $\psi_u$ (the true quality for discrete model) for $\sigma=0.1$. The straight line represents an ideal model. The red line is the obtained value.}\label{fig:smallvar}
\end{figure}

\begin{figure}
	\centering
	\input{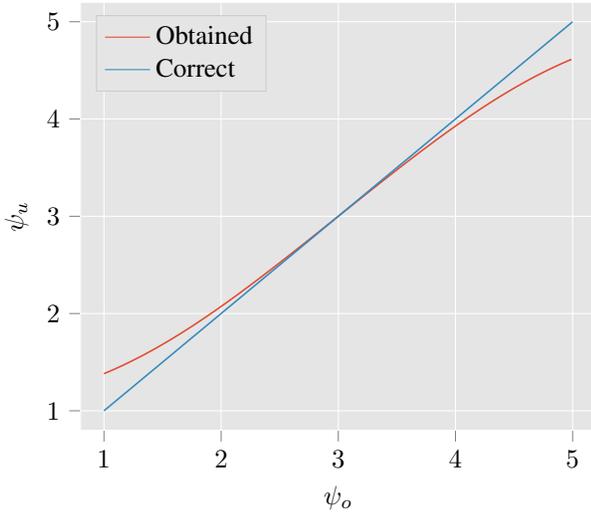}
	\caption{The difference between $\psi_{o}$ (the true quality for continuous model) and $\psi_{u}$ (the true quality for discrete model) for $\sigma=1$. The straight line represents an ideal model. The red line is the obtained value.}\label{fig:largevar}
\end{figure}

Note that the parameters estimated from $O$ and $U$ can result in different $\psi$ and $\sigma$ values since the support of $O$ is any real number and the support of $U$ is set $\{1, 2, \cdots, 5\}$. Let us mark by $\psi_o$ and $\sigma_o$ the parameters of $O$ distribution which are our input parameters and by $\psi_u$ and $\sigma_u$ the mean and standard deviation of random variable $U$. For a given input pair $(\psi_o, \sigma_o)$ the comparison with the obtained $(\psi_u, \sigma_u)$ is crucial, since we interpret the input parameters but the real process will behave according to the output parameters. For the ideal model those two values should be identical. In Fig.~\ref{fig:smallvar}~and~\ref{fig:largevar} we compared the relation between $\psi_o$ and $\psi_u$ for $\sigma_o = 0.1$ and $\sigma_o=1$ respectively.

Both Fig.~\ref{fig:smallvar}~and~\ref{fig:largevar} show large differences between $\psi_o$ and $\psi_u$. To understand this discrepancy we have to understand the dissimilarities of a continuous and discrete process. For continuous process the parameters can have any possible value so $\psi_o \in (-\infty, \infty)$ and $\sigma_o \in [0, \infty)$. On the other hand, $\psi_u \in [1, 5]$ and the range of $\sigma_u$ depends on $\psi_u$. This is very different from the continuous model, where those parameters are independent. This dependency is caused by a finite number of arguments values (i.e., subject answers) of the discrete distributions. For example, if $\psi=1.1$ the two most extreme, in terms of variance, parameters are:
\begin{enumerate}
	\item $P(U=1)=0.9, P(U=2)=0.1$ with $\sigma^2 = 0.09$
	\item $P(U=1)=0.975, P(U=5)=0.025$ with $\sigma^2 = 0.39$.
\end{enumerate}
On the other hand, if $\psi=3$ the two most extreme distributions are:
\begin{enumerate}
	\item $P(U=3)=1$ with $\sigma^2 = 0.0$
	\item $P(U=1)=0.5, P(U=5)=0.5$ with $\sigma^2 = 4$.
\end{enumerate}

\begin{figure}
	\centering
	\includegraphics{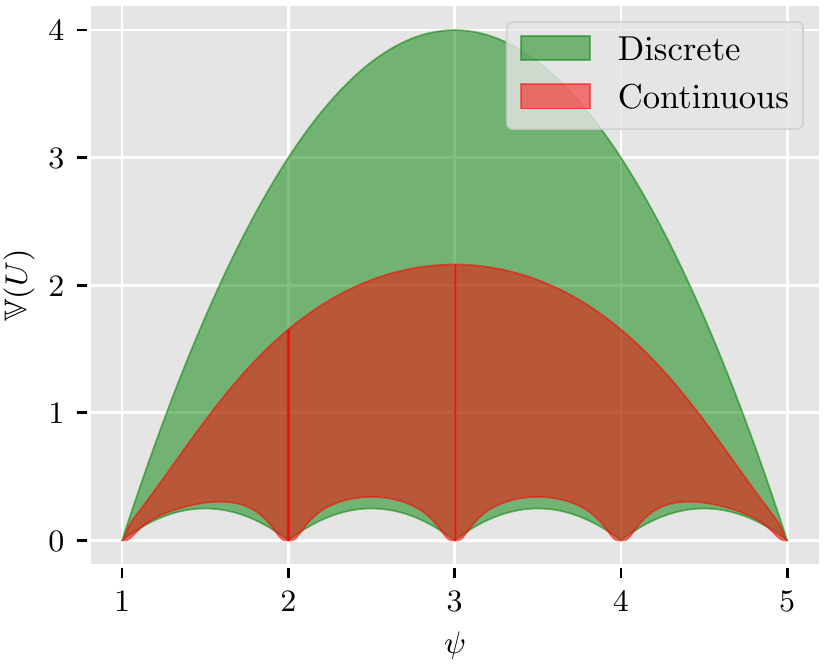}
	\caption{The limitation of $\mathbb{V}(U)$ as a function of $\psi$. Green surface shows the limitation for discrete distribution. Red surface is limitation of $\sigma_u^2$ for the continuous model if the discrete distribution limitations are used as an input $\sigma_o^2$.}\label{fig:varLimit}
\end{figure}

The above examples show that the limited support of the discrete distribution limits the possible value of $\sigma$ in a complicated way. The full dependency is shown in Fig.~\ref{fig:varLimit} and marked by the green surface. More details about the variance limitation are in \cite{Hossfeld2011}\footnote{Note that Fig.~\ref{fig:varLimit} in \cite{Hossfeld2011} is presented for standard deviation and not variance. Therefore, it looks different, but represents the same phenomenon.}.

In order to use the continuous model we typically limit the parameters of the model to the range of the discrete parameters. This solution works well for $\psi_o$, which can be limited to $[1,5]$ interval and so the $\psi_u$ will be limited in the same, correct interval. Nevertheless, if we limit $\sigma_o^2$ to the interval defined by the green surface in Fig.~\ref{fig:varLimit}, the obtained $\sigma_u^2$ is given by the red surface. As we can see there is no match between $\sigma_o^2$ and $\sigma_u^2$. This is a strong limitation of the continuous model and cannot be easily solved. Therefore, we propose the following discrete distribution. 

\subsection{Discrete Distribution}

An alternative solution to a continuous model is the discrete distribution\footnote{We use the term \textit{discrete distribution} (instead of \textit{discrete model}) purposely. When using a continuous distribution we need to use a model to link it to discrete values. However, when modeling discrete values by a discrete distribution we can think of using this particular distribution as a model.}. By the discrete distribution we understand the specific discrete distribution for which a probability of each answer is given by a function. We are looking for a function of two parameters, allowing us to specify true quality $\psi$ and the answers' precision/spread $\sigma$. Since the discrete distribution parameter describing the distribution variance is not the same as the process variance we mark it by $\rho$ and not $\sigma$. 

According to our knowledge there is no distribution described in literature which is:
\begin{itemize}
	\item discrete
	\item supported on finite $M$ element set 
	\item for any fixed mean covers the whole spectrum of possible variances
	\item with no more than a single change in the probabilities monotonicity. 
\end{itemize} 
Therefore, we propose a new distribution with two parameters describing mean $\psi$ and variance $\rho$. The $\rho$ parameter describes variance for a given $\psi$. Since the proposed distribution is discrete the problems shown on Fig.~\ref{fig:smallvar}~and~\ref{fig:largevar} do not exist. 

The proposed distribution utilizes existing discrete distributions (with limited support), but covers more possible variances. An example of a discrete distribution which is described by two parameters is Beta Binomial distribution \cite{betaBinomial}. Nevertheless, the smallest variance of the Beta Binomial distribution is the Binomial distribution variance. It is a strong limitation (in the case of subjective experiments for video) since in \cite{Hossfeld2018} it is suggested that the Binomial distribution has the highest possible variance for a correctly conducted subjective experiment. Therefore, we need a different distribution. 

\subsubsection{GSD Definition}

Let us start with the equation describing the answers $U$ for a particular PVS
\begin{equation}
U = \psi + \epsilon,
\end{equation}
where $\epsilon$ is an error with mean value equals to $0$. Since $U$ belongs to the set $\{1,2,...,5\}$ then the distribution of $\epsilon$ has to be supported on the set $1-\psi$, $2-\psi$, ..., $5-\psi$. 
Let us consider shifted Binomial distribution for $\epsilon$:
$$P(\epsilon=k-\psi)=\binom{4}{k-1}\left(\frac{\psi-1}{4}\right)^{k-1}\left(\frac{5-\psi}{4}\right)^{5-k},$$
where $k\in\{1,...,5\}$ is an user answer. 

Since the support of this distribution and the mean value are fixed, we obtain fixed shifted Binomial distribution without any freedom. However, we would like to have a class of distributions for $\epsilon$ with all possible variances. Let us think about how the set of all possible variances, for all distributions supported on the set $1-\psi$, $2-\psi$, ..., $5-\psi$ looks like. Remember that the mean values for such distributions are fixed to $0$. This is why the set of all possible variances depends on $\psi$. If we denote by $V_{\mathrm{min}}(\psi), V_{\mathrm{max}}(\psi)$ the minimal and maximal possible variance respectively, then
$$V_{\mathrm{min}}(\psi)=(\lceil\psi\rceil-\psi)(\psi-\lfloor\psi\rfloor),$$
$$V_{\mathrm{max}}(\psi)=(\psi-1)(5-\psi),$$
and the interval $[V_{\mathrm{min}}(\psi),V_{\mathrm{max}}(\psi)]$ is the set of all possible variances. Notice that the interval $[V_{\mathrm{min}}(\psi),V_{\mathrm{max}}(\psi)]$ is the biggest for $\psi=3$ which is $[V_{\mathrm{min}}(3),V_{\mathrm{max}}(3)]=[0,4]$, and if $\psi$ is not an integer then $V_{\mathrm{min}}(\psi)>0$.
Let us return to the shifted Binomial distribution. It is easy to calculate that its variance is equal to: 
\begin{equation}\label{eq:VarBin}
V_{\mathrm{Bin}}(\psi):=\frac{V_{\mathrm{max}}(\psi)}{4}.
\end{equation}
The question is how to obtain from this shifted Binomial distribution a class of distributions that covers the whole interval of variances $[V_{\mathrm{min}}(\psi),V_{\mathrm{max}}(\psi)]$ (see Fig. \ref{fig:varRel}). 

\begin{figure}
	\centering
	\begin{tikzpicture}[xscale=8]
	\draw (0, 0.5) node{$\psi=1.3$};
	\draw (0.14725, 0.5) node{$\circ$};
	\draw (0.1624375, 0.5) node{$\times$};
	\draw (0.34975, 0.5) node{$\bullet$};
	\draw [-][draw=red, very thick] (0.14725, 0.5) -- (0.1624375, 0.5);
	\draw [-] [draw = green, very thick](0.1624375, 0.5) -- (0.34975, 0.5);
	
	\draw (0, 0) node{$\psi=2.6$};
	\draw (0.154, 0) node{$\circ$};
	\draw (0.316, 0) node{$\times$};
	\draw (0.964, 0) node{$\bullet$};
	\draw [-][draw=red, very thick] (0.154, 0) -- (0.316, 0);
	\draw [-] [draw = green, very thick](0.316, 0) -- (0.964, 0);
	
	\draw (0,-0.5) node{$\psi=4.3$};
	\draw (0.14725, -0.5) node{$\circ$};
	\draw (0.2299375, -0.5) node{$\times$};
	\draw (0.61975, -0.5) node{$\bullet$};
	\draw [-][draw=red, very thick] (0.14725, -0.5) -- (0.2299375, -0.5);
	\draw [-] [draw = green, very thick](0.2299375, -0.5) -- (0.61975, -0.5);
	
	\draw (0,-1) node{$\psi=3.0$};
	\draw[-][draw=red, very thick] (0.1,-1) -- (0.325,-1);
	\draw[-][draw=green, very thick] (0.325,-1) -- (1,-1);
	\draw (0.1, -1) node{$\circ$};
	\draw (0.325, -1) node{$\times$};
	\draw (1, -1) node{$\bullet$};
	\draw [thick] (0.1,-1.1) node[below]{0} -- (0.1,-0.9);
	\draw [thick] (0.325,-1.1) node[below]{1} -- (0.325,-0.9);
	\draw [thick] (1,-1.1) node[below]{4} -- (1,-0.9);
	
	\draw (0, -2) node{Legend:};
	\draw (0.1, -2) node{$\circ$};
	\draw (0.20, -2) node{$V_{\mathrm{min}}(\psi)$,};
	\draw (0.32, -2) node{$\times$};
	\draw (0.42, -2) node{$V_{\mathrm{Bin}}(\psi)$,};
	\draw (0.54, -2) node{$\bullet$};
	\draw (0.64, -2) node{$V_{\mathrm{max}}(\psi)$};
	\end{tikzpicture}
	
	\caption{Visualization of the the interval $[V_{\mathrm{min}}(\psi),V_{\mathrm{max}}(\psi)]$ for different $\psi$ values. The interval $[V_{\mathrm{min}}(\psi),V_{\mathrm{max}}(\psi)]$ is the biggest for $\psi=3$ and has the form of $[V_{\mathrm{min}}(3),V_{\mathrm{max}}(3)]=[0,4]$. The red and green intervals are the possible variances, smaller and grater than $V_{\mathrm{Bin}}(\psi)$ respectively.}\label{fig:varRel}
\end{figure}

\begin{figure}
	\begin{center}
		\input{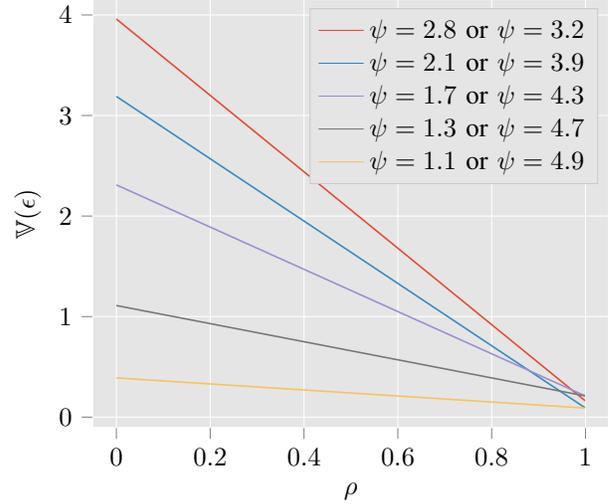}
		\caption{Variances of the error term $\epsilon$} \label{fig:varMOS}
	\end{center}
\end{figure}

We would like to obtain this class by:
\begin{itemize}
	\item adding only a single normalized parameter $\rho\in(0,1]$, 
	\item variance of the error to be linearly dependent on $\rho$,
	\item variance is a decreasing function of $\rho$. 
\end{itemize}
In this way, we would consider the $\rho$ parameter a confidence parameter (see Fig. \ref{fig:varMOS}).

Let us denote by $H_{\rho}$ the distribution of the error fulfilling the above conditions. Since the variance of the error is linearly dependent on $\rho$ and decreasing then it has to be equal to
\begin{equation}\label{eq:Var}
\mathbb{V}_{H_{\rho}}(\epsilon)=\rho V_{\mathrm{min}}(\psi)+(1-\rho)V_{\mathrm{max}}(\psi).
\end{equation}  
Using formulas (\ref{eq:VarBin}) and (\ref{eq:Var}) we can easily calculate 
$$
\mathbb{V}_{H_{\rho}}(\epsilon)=V_{\mathrm{Bin}}(\psi) \  \Rightarrow  \ \rho=C(\psi):=\frac{3}{4}\ \frac{V_{\mathrm{max}}(\psi)}{V_{\mathrm{max}}(\psi)-V_{\mathrm{min}}(\psi)},
$$ 
which gives us the value of $\rho$ corresponding to shifted binomial distribution. We have 
$$
\mathbb{V}_{H_{\rho}}(\epsilon)\in [V_{\mathrm{min}}(\psi),V_{\mathrm{Bin}}(\psi)] \Leftrightarrow \rho \in [C(\psi),1],
$$
which corresponds to the red coloured  interval in Fig. \ref{fig:varRel}, and
$$
\mathbb{V}_{H_{\rho}}(\epsilon)\in [V_{\mathrm{Bin}}(\psi),V_{\mathrm{max}}(\psi)] \Leftrightarrow \rho \in [0,C(\psi)],
$$
which corresponds to the green coloured interval in Fig. \ref{fig:varRel}. 

For variances bigger than $V_{\mathrm{Bin}}(\psi)$ (the green coloured interval in Fig. \ref{fig:varRel}) we use the Beta Binomial distribution. In other words, we replace success probability parameter by a random variable with the Beta distribution. Since the mean value is fixed, we only have one free parameter. It can be reparameterized to the interval $(0,C(\psi))$. The effect of such reparameterization gives us the distribution denoted by $G_{\rho}$:
\begin{equation}
\begin{split}
P_{G_{\rho}}&(\epsilon=k-\psi)= \binom{4}{k-1} \times \\
&  \frac{\mathcal{B}\left(\frac{\rho(\psi-1)}{4(C(\psi)-\rho)}+k-1,\frac{(5-\psi)\rho}{4(C(\psi)-\rho)}+5-k\right)}{\mathcal{B}\left(\frac{(\psi-1)\rho}{4(C(\psi)-\rho)},\frac{(5-\psi)\rho}{4(C(\psi)-\rho)}\right)}, \\
\end{split} \label{eq:Prho}
\end{equation}
where $\rho\in(0,C(\psi))$ and $k\in\{1,...,5\}$. 

\begin{re}
	Notice that for $\rho \rightarrow 0$ the distribution $G_{\rho}$ goes to a two points distribution supported on $\{1-\psi,5-\psi\}$ with the biggest possible variance equal to $V_{\mathrm{max}}(\psi)$. For $\rho \rightarrow C(\psi)$ the distribution $G_{\rho}$ goes to the shifted Binomial distribution with variance equal to $V_{\mathrm{Bin}}(\psi)$.
\end{re}

For variances smaller than $V_{\mathrm{Bin}}(\psi)$ (the red coloured interval in Fig. \ref{fig:varRel}) we use a mixture technique. Specifically, we take a mixture of the shifted Binomial distribution with the distribution with the smallest possible variance (two points or one point distribution, depending on $\psi$). Of course the mixture parameter has to be reparameterized to the interval $[C(\psi),1]$.
The effect of such reparameterization gives us the distribution denoted by $F_{\rho}$:
\begin{equation} 
\begin{split}
P_{F_{\rho}} & (\epsilon = k-\psi)= \\ 
& \frac{\rho-C(\psi)}{1-C(\psi)} [1-|k-\psi|]_{+} + \\
& \frac{1-\rho}{1-C(\psi)}\binom{4}{k-1}\left(\frac{\psi-1}{4}\right)^{k-1}\left(\frac{5-\psi}{4}\right)^{5-k}, \\
\end{split} \label{eq:Frho}
\end{equation}
where $\rho\in[C(\psi),1]$, $[x]_{+} = \max(x,0)$ and $k\in\{1,...,5\}$.

\begin{re}
	Notice that for $\rho \rightarrow C(\psi)$ the distribution $F_{\rho}$ goes to the shifted Binomial distribution with variance equal to $V_{\mathrm{Bin}}(\psi)$.
	For $\rho \rightarrow 1$ the distribution $F_{\rho}$ goes to a two points or one point distribution (depending on $\psi$) with the smallest possible variance equal to $V_{\mathrm{min}}(\psi)$.
\end{re}

Finally, we obtain the desired GSD:
$$H_{\rho}=G_{\rho}\ I(\rho<C(\psi)) + F_{\rho}\ I(\rho\geq C(\psi)),$$
where $\rho\in(0,1]$ is a confidence parameter and $I(x)$ is one if $x$ is true or 0 if $x$ is false. $I(x)$ can be seen as an \textit{if} function. The variance in that distribution fulfills equation (\ref{eq:Var}).

In order to describe subjective scores correctly and completely, a process covering the whole spectrum of variances is needed. Our discrete distribution $H_{\rho}$ addresses this claim. Furthermore, we believe the proposed distribution can be useful for describing other, more general processes. 

From this moment we will assume the following
\begin{as}\label{eq:mod} 
	$$U=\psi+\epsilon,$$
	where $\psi\in[1,5]$ is an unknown parameter and $\epsilon$ is a random variable with distribution functions $H_{\rho}$, where $\rho\in(0,1]$ is an unknown parameter.
\end{as}

Examples of the $U$ distributions are shown in Fig.~\ref{fig:disExa}. Note that for $\psi$ close to 1 or 5, regardless of $\rho$, the obtained distributions are similar. It comes from the fact, that the maximum spread we can obtain, is limited by a small range of possible variances (see Fig.~\ref{fig:varLimit}). For $\psi$ closer to 3 the obtained distributions are clearly different.

\afterpage{%
	\begin{figure*}
		\begin{center}
			\subfloat{
\begin{tikzpicture}

\definecolor{color0}{rgb}{0.886274509803922,0.290196078431373,0.2}
\definecolor{color1}{rgb}{0.203921568627451,0.541176470588235,0.741176470588235}
\definecolor{color2}{rgb}{0.596078431372549,0.556862745098039,0.835294117647059}
\definecolor{color3}{rgb}{0.984313725490196,0.756862745098039,0.368627450980392}
\definecolor{color4}{rgb}{0.556862745098039,0.729411764705882,0.258823529411765}
\definecolor{color5}{rgb}{0.75,0,0.75}

\begin{axis}[
axis background/.style={fill=white!89.80392156862746!black},
axis line style={white},
legend cell align={left},
legend entries={{0.95},{0.88},{0.81},{0.72},{0.61},{0.38}},
legend style={draw=white!80.0!black, fill=white!89.80392156862746!black},
tick align=outside,
tick pos=left,
x grid style={white},
xlabel={Score $s$},
xmajorgrids,
xmin=0.8, xmax=5.2,
y grid style={white},
ylabel={$P(U_{ij} = s)$},
ymajorgrids,
ymin=-0.05, ymax=1.05
]
\addplot [line width=0.08000000000000002pt, color0, dashed, mark=*, mark size=4, mark options={solid}]
table [row sep=\\]{%
1	0.72139609375 \\
2	0.258290625 \\
3	0.0192515625 \\
4	0.001040625 \\
5	2.109375e-05 \\
};
\addplot [line width=0.08000000000000002pt, color1, dashed, mark=triangle*, mark size=4, mark options={solid,rotate=180}]
table [row sep=\\]{%
1	0.748496941750038 \\
2	0.208222542278198 \\
3	0.038350722665177 \\
4	0.00464316083489919 \\
5	0.000286632471687573 \\
};
\addplot [line width=0.08000000000000002pt, color2, dashed, mark=diamond*, mark size=4, mark options={solid}]
table [row sep=\\]{%
1	0.77096417921922 \\
2	0.171207946584666 \\
3	0.0460910849306944 \\
4	0.0103372735077423 \\
5	0.00139951575767982 \\
};
\addplot [line width=0.08000000000000002pt, white!46.666666666666664!black, dashed, mark=x, mark size=4, mark options={solid}]
table [row sep=\\]{%
1	0.795829157800537 \\
2	0.134191333399622 \\
3	0.0484510529979119 \\
4	0.017207262603162 \\
5	0.00432119319876703 \\
};
\addplot [line width=0.08000000000000002pt, color3, dashed, mark=pentagon*, mark size=4, mark options={solid}]
table [row sep=\\]{%
1	0.821895746624118 \\
2	0.0996567312454621 \\
3	0.0451553265189036 \\
4	0.0231361667293331 \\
5	0.0101560288821829 \\
};
\addplot [line width=0.08000000000000002pt, color4, dashed, mark=square*, mark size=4, mark options={solid}]
table [row sep=\\]{%
1	0.866684689952905 \\
2	0.0497367503924647 \\
3	0.0296816091051805 \\
4	0.0246877708006279 \\
5	0.0292091797488226 \\
};
\addplot [very thick, color5, forget plot]
table [row sep=\\]{%
1.3	0 \\
1.3	1 \\
};
\node at (axis cs:1.4,0.8)[
  anchor=base west,
  text=black,
  rotate=0.0
]{ $\psi = 1.30$};
\end{axis}

\end{tikzpicture}}
			\subfloat{
\begin{tikzpicture}

\definecolor{color0}{rgb}{0.886274509803922,0.290196078431373,0.2}
\definecolor{color1}{rgb}{0.203921568627451,0.541176470588235,0.741176470588235}
\definecolor{color2}{rgb}{0.596078431372549,0.556862745098039,0.835294117647059}
\definecolor{color3}{rgb}{0.984313725490196,0.756862745098039,0.368627450980392}
\definecolor{color4}{rgb}{0.556862745098039,0.729411764705882,0.258823529411765}
\definecolor{color5}{rgb}{0.75,0,0.75}

\begin{axis}[
axis background/.style={fill=white!89.80392156862746!black},
axis line style={white},
legend cell align={left},
legend entries={{0.95},{0.88},{0.81},{0.72},{0.61},{0.38}},
legend style={draw=white!80.0!black, fill=white!89.80392156862746!black},
tick align=outside,
tick pos=left,
x grid style={white},
xlabel={Score $s$},
xmajorgrids,
xmin=0.8, xmax=5.2,
y grid style={white},
ylabel={$P(U_{ij} = s)$},
ymajorgrids,
ymin=-0.05, ymax=1.05
]
\addplot [line width=0.08000000000000002pt, color0, dashed, mark=*, mark size=4, mark options={solid}]
table [row sep=\\]{%
1	0.0605281332818022 \\
2	0.794662643551237 \\
3	0.130343269655477 \\
4	0.0132129969081272 \\
5	0.00125295660335689 \\
};
\addplot [line width=0.08000000000000002pt, color1, dashed, mark=triangle*, mark size=4, mark options={solid,rotate=180}]
table [row sep=\\]{%
1	0.145267519876325 \\
2	0.647190344522968 \\
3	0.172823847173145 \\
4	0.0317111925795053 \\
5	0.00300709584805654 \\
};
\addplot [line width=0.08000000000000002pt, color2, dashed, mark=diamond*, mark size=4, mark options={solid}]
table [row sep=\\]{%
1	0.230006906470848 \\
2	0.4997180454947 \\
3	0.215304424690813 \\
4	0.0502093882508834 \\
5	0.00476123509275618 \\
};
\addplot [line width=0.08000000000000002pt, white!46.666666666666664!black, dashed, mark=x, mark size=4, mark options={solid}]
table [row sep=\\]{%
1	0.316802044146678 \\
2	0.37043976700906 \\
3	0.221868434092756 \\
4	0.0777356542005991 \\
5	0.0131541005509078 \\
};
\addplot [line width=0.08000000000000002pt, color3, dashed, mark=pentagon*, mark size=4, mark options={solid}]
table [row sep=\\]{%
1	0.394171811991077 \\
2	0.285141337039146 \\
3	0.184045116936097 \\
4	0.0997985070460573 \\
5	0.0368432269876218 \\
};
\addplot [line width=0.08000000000000002pt, color4, dashed, mark=square*, mark size=4, mark options={solid}]
table [row sep=\\]{%
1	0.532202936822452 \\
2	0.153334274334812 \\
3	0.107779556060854 \\
4	0.0956263175840495 \\
5	0.111056915197833 \\
};
\addplot [very thick, color5, forget plot]
table [row sep=\\]{%
2.1	0 \\
2.1	1 \\
};
\node at (axis cs:2.2,0.8)[
  anchor=base west,
  text=black,
  rotate=0.0
]{ $\psi = 2.10$};
\end{axis}

\end{tikzpicture}}
			\\
			\vspace*{-0.3cm}
			\subfloat{
\begin{tikzpicture}

\definecolor{color0}{rgb}{0.886274509803922,0.290196078431373,0.2}
\definecolor{color1}{rgb}{0.203921568627451,0.541176470588235,0.741176470588235}
\definecolor{color2}{rgb}{0.596078431372549,0.556862745098039,0.835294117647059}
\definecolor{color3}{rgb}{0.984313725490196,0.756862745098039,0.368627450980392}
\definecolor{color4}{rgb}{0.556862745098039,0.729411764705882,0.258823529411765}
\definecolor{color5}{rgb}{0.75,0,0.75}

\begin{axis}[
axis background/.style={fill=white!89.80392156862746!black},
axis line style={white},
legend cell align={left},
legend entries={{0.95},{0.88},{0.81},{0.72},{0.61},{0.38}},
legend style={draw=white!80.0!black, fill=white!89.80392156862746!black},
tick align=outside,
tick pos=left,
x grid style={white},
xlabel={Score $s$},
xmajorgrids,
xmin=0.8, xmax=5.2,
y grid style={white},
ylabel={$P(U_{ij} = s)$},
ymajorgrids,
ymin=-0.05, ymax=1.05
]
\addplot [line width=0.08000000000000002pt, color0, dashed, mark=*, mark size=4, mark options={solid}]
table [row sep=\\]{%
1	0.0185348096301595 \\
2	0.18048492784562 \\
3	0.743586358682183 \\
4	0.0472332605781363 \\
5	0.0101606432639014 \\
};
\addplot [line width=0.08000000000000002pt, color1, dashed, mark=triangle*, mark size=4, mark options={solid,rotate=180}]
table [row sep=\\]{%
1	0.0444835431123828 \\
2	0.223163826829488 \\
3	0.594607260837239 \\
4	0.113359825387527 \\
5	0.0243855438333634 \\
};
\addplot [line width=0.08000000000000002pt, color2, dashed, mark=diamond*, mark size=4, mark options={solid}]
table [row sep=\\]{%
1	0.0704322765946062 \\
2	0.265842725813356 \\
3	0.445628162992295 \\
4	0.179486390196918 \\
5	0.0386104444028254 \\
};
\addplot [line width=0.08000000000000002pt, white!46.666666666666664!black, dashed, mark=x, mark size=4, mark options={solid}]
table [row sep=\\]{%
1	0.114405372370277 \\
2	0.276308916717531 \\
3	0.323641015625745 \\
4	0.216169729114808 \\
5	0.0694749661716385 \\
};
\addplot [line width=0.08000000000000002pt, color3, dashed, mark=pentagon*, mark size=4, mark options={solid}]
table [row sep=\\]{%
1	0.178081405530565 \\
2	0.244367463186289 \\
3	0.249159177257745 \\
4	0.206253633803385 \\
5	0.122138320222017 \\
};
\addplot [line width=0.08000000000000002pt, color4, dashed, mark=square*, mark size=4, mark options={solid}]
table [row sep=\\]{%
1	0.313469658678537 \\
2	0.158680186692613 \\
3	0.136641487850938 \\
4	0.146797829506136 \\
5	0.244410837271775 \\
};
\addplot [very thick, color5, forget plot]
table [row sep=\\]{%
2.85	0 \\
2.85	1 \\
};
\node at (axis cs:2.95,0.8)[
  anchor=base west,
  text=black,
  rotate=0.0
]{ $\psi = 2.85$};
\end{axis}

\end{tikzpicture}}
			\subfloat{
\begin{tikzpicture}

\definecolor{color0}{rgb}{0.886274509803922,0.290196078431373,0.2}
\definecolor{color1}{rgb}{0.203921568627451,0.541176470588235,0.741176470588235}
\definecolor{color2}{rgb}{0.596078431372549,0.556862745098039,0.835294117647059}
\definecolor{color3}{rgb}{0.984313725490196,0.756862745098039,0.368627450980392}
\definecolor{color4}{rgb}{0.556862745098039,0.729411764705882,0.258823529411765}
\definecolor{color5}{rgb}{0.75,0,0.75}

\begin{axis}[
axis background/.style={fill=white!89.80392156862746!black},
axis line style={white},
legend cell align={left},
legend entries={{0.95},{0.88},{0.81},{0.72},{0.61},{0.38}},
legend style={at={(0.03,0.97)}, anchor=north west, draw=white!80.0!black, fill=white!89.80392156862746!black},
tick align=outside,
tick pos=left,
x grid style={white},
xlabel={Score $s$},
xmajorgrids,
xmin=0.8, xmax=5.2,
y grid style={white},
ylabel={$P(U_{ij} = s)$},
ymajorgrids,
ymin=-0.05, ymax=1.05
]
\addplot [line width=0.08000000000000002pt, color0, dashed, mark=*, mark size=4, mark options={solid}]
table [row sep=\\]{%
1	0.00125295660335689 \\
2	0.0132129969081272 \\
3	0.130343269655477 \\
4	0.794662643551237 \\
5	0.0605281332818022 \\
};
\addplot [line width=0.08000000000000002pt, color1, dashed, mark=triangle*, mark size=4, mark options={solid,rotate=180}]
table [row sep=\\]{%
1	0.00300709584805654 \\
2	0.0317111925795053 \\
3	0.172823847173145 \\
4	0.647190344522968 \\
5	0.145267519876325 \\
};
\addplot [line width=0.08000000000000002pt, color2, dashed, mark=diamond*, mark size=4, mark options={solid}]
table [row sep=\\]{%
1	0.00476123509275618 \\
2	0.0502093882508834 \\
3	0.215304424690813 \\
4	0.4997180454947 \\
5	0.230006906470848 \\
};
\addplot [line width=0.08000000000000002pt, white!46.666666666666664!black, dashed, mark=x, mark size=4, mark options={solid}]
table [row sep=\\]{%
1	0.0131541005509078 \\
2	0.077735654200599 \\
3	0.221868434092756 \\
4	0.370439767009059 \\
5	0.316802044146677 \\
};
\addplot [line width=0.08000000000000002pt, color3, dashed, mark=pentagon*, mark size=4, mark options={solid}]
table [row sep=\\]{%
1	0.0368432269876218 \\
2	0.0997985070460573 \\
3	0.184045116936097 \\
4	0.285141337039146 \\
5	0.394171811991077 \\
};
\addplot [line width=0.08000000000000002pt, color4, dashed, mark=square*, mark size=4, mark options={solid}]
table [row sep=\\]{%
1	0.111056915197833 \\
2	0.0956263175840493 \\
3	0.107779556060854 \\
4	0.153334274334812 \\
5	0.532202936822451 \\
};
\addplot [very thick, color5, forget plot]
table [row sep=\\]{%
3.9	0 \\
3.9	1 \\
};
\node at (axis cs:2.9,0.8)[
  anchor=base west,
  text=black,
  rotate=0.0
]{ $\psi = 3.90$};
\end{axis}

\end{tikzpicture}}
			\caption{Distributions of $U$ under Assumption \ref{eq:mod} for different $\psi$ and $\rho$.} \label{fig:disExa}
		\end{center}
	\end{figure*}
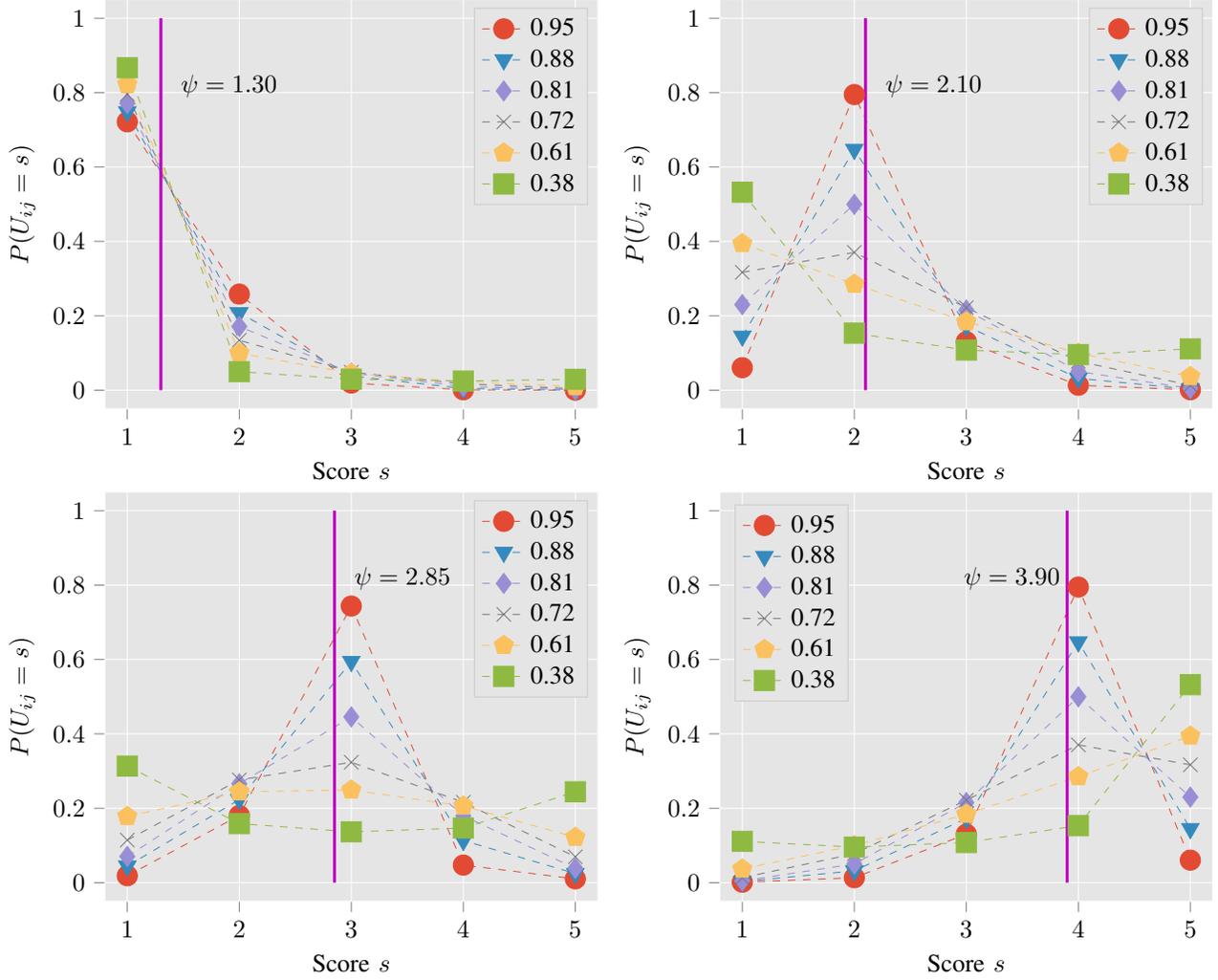
	
}

\section{GSD Parameters Estimation}

\label{sec:estimation}

Continuous model and discrete distribution described in the previous sections cannot be used without an accurate and efficient parameter estimation procedure. The most simple approach is to use method of moments. However, in the numerical experiments, we use a much better Maximum Likelihood Estimator (MLE). The distribution of $U$ is parameterized by a pair $\theta = (\psi,\rho)$ in the case of discrete distribution or $\theta = (\psi,\sigma)$ for the continuous model. Let $(u_{i})$ be a vector of iid samples from $U$.

An MLE is defined as:
\begin{equation}
\hat \theta = \arg\max_\theta \ell(\mathbf u|\theta),
\end{equation}
where 
\begin{equation}\label{eq:ll}
\ell(\mathbf u|\theta) =  \sum_i \ln P_{\theta}(U=u_{i})
\end{equation}
is the likelihood function of distribution $U$. Since there is no closed form for a maximum of $\ell(\theta)$, a numerical optimization is used to maximize the equation~\eqref{eq:ll}. 
The objective function~\eqref{eq:ll} is maximized using RMSProp, a first order gradient-based optimizer. 
Optimization stopping criteria are 
\begin{equation*}
\lvert \nabla \ell(\theta) \rvert \leq \epsilon
\end{equation*}
or the number of steps reached some fixed value -- whatever is first.

Since the paper proposes a new distribution, we have to implement the likelihood function from scratch. 
The logarithm of \eqref{eq:Frho} and \eqref{eq:Prho} is implemented using low-level TensorFlow ops (i.e., operations), which allows us to take advantage of efficient implementations of special function and automatic differentiation for gradient-based optimizers provided by the library. Details about the implementation are in Appendix~\ref{sec:ap:est}.

\subsection{Numerical Experiments}
In order to validate our estimation procedures we drawn data from the proposed distribution and estimate the distribution parameters from the sampled data. The simulation considers:
\begin{itemize}
	\item a number of subjects $N = \left\{6, 12, 24, 48\right\}$,
	\item $\psi$ from 1.05 to 4.95 (with 23 different values),
	\item $\rho$ from 0.01 to 0.99 (with 23 different values).
\end{itemize}

For each triple ($\psi$, $\rho$, and $N$) the simulation is repeated 30 times.

For 32 cases we obtain a fast convergence (less than 1,000 steps). For 6 points the estimated parameters are clear outliers. We assume those are just numerical errors and for experiment data they are easy to avoid by estimating the parameters more than one time and choosing the result with lower likelihood value. 

\begin{figure}
	\begin{center}
		\input{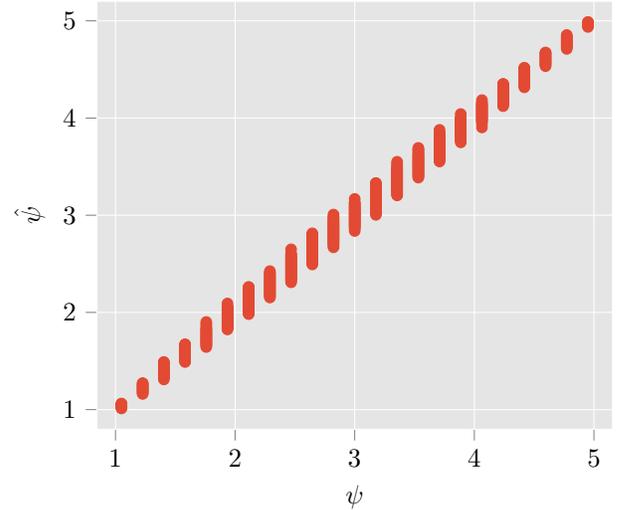}
		\caption{Comparison between $\psi$ as a simulation parameter and the estimated value $\hat{\psi}$ for all simulations.} \label{fig:psi}
	\end{center}
\end{figure}

\begin{figure}
	\begin{center}
		\input{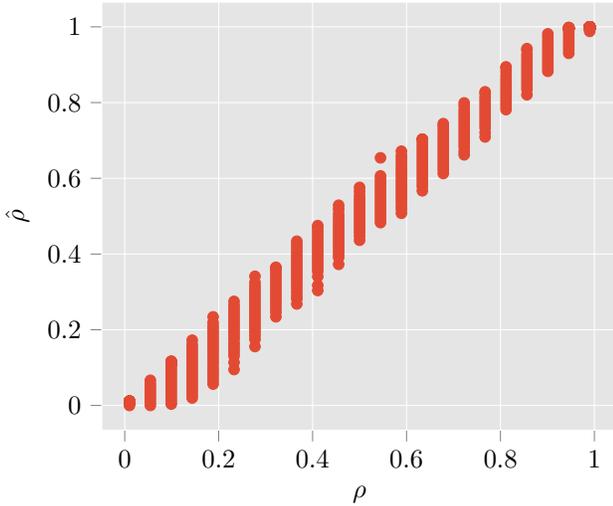}
		\caption{Comparison between $\rho$ as a simulation parameter and the estimated value $\hat{\rho}$ for all simulations.} \label{fig:rho}
	\end{center}
\end{figure}

After removing the 6 numerically unstable points we obtain a high accuracy (see Fig.~\ref{fig:psi} and~\ref{fig:rho}). The true quality $\psi$ is typically in the range $\pm0.25$ and $\rho$ is in the range $\pm 0.1$. This is including simulations with only 6 answers.

\begin{figure*}
	\begin{center}
		\pgfplotsset{every axis title/.append style={at={(0.15,0.8)}}}
		\input{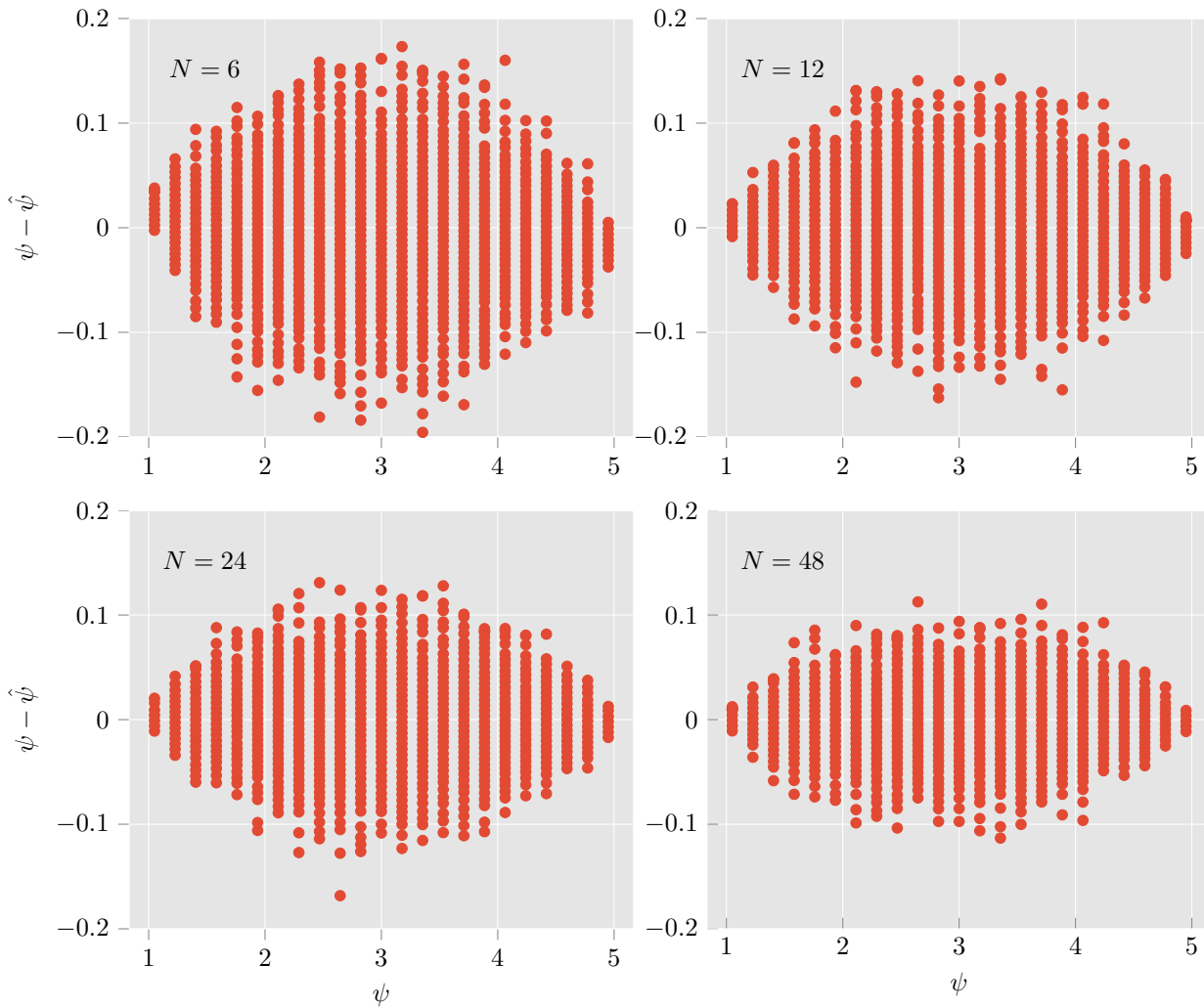}
		\caption{Comparison between $\psi$ and the estimation error calculated as $\psi - \hat{\psi}$ for different number of subjects $N$.} \label{fig:psi_per_n}
	\end{center}
\end{figure*}

To better analyze the obtained accuracy Fig.~\ref{fig:psi_per_n} shows the accuracy measured by $\psi - \hat{\psi}$ for different number of subjects. As expected, using a higher number of subjects increases accuracy. We can also see clearly the estimation closer to the edges is less scattered. Another conclusion is the confidence interval for $\psi$ should be close to $\pm0.2$ for $N=6$ and $\pm 0.1$ for $N=24$. Of course, more formal analysis is needed to obtain correct confidence interval estimation. This problem is left as future work.

\begin{figure*}
	\begin{center}
		\pgfplotsset{every axis title/.append style={at={(0.85,0.8)}}}
		\input{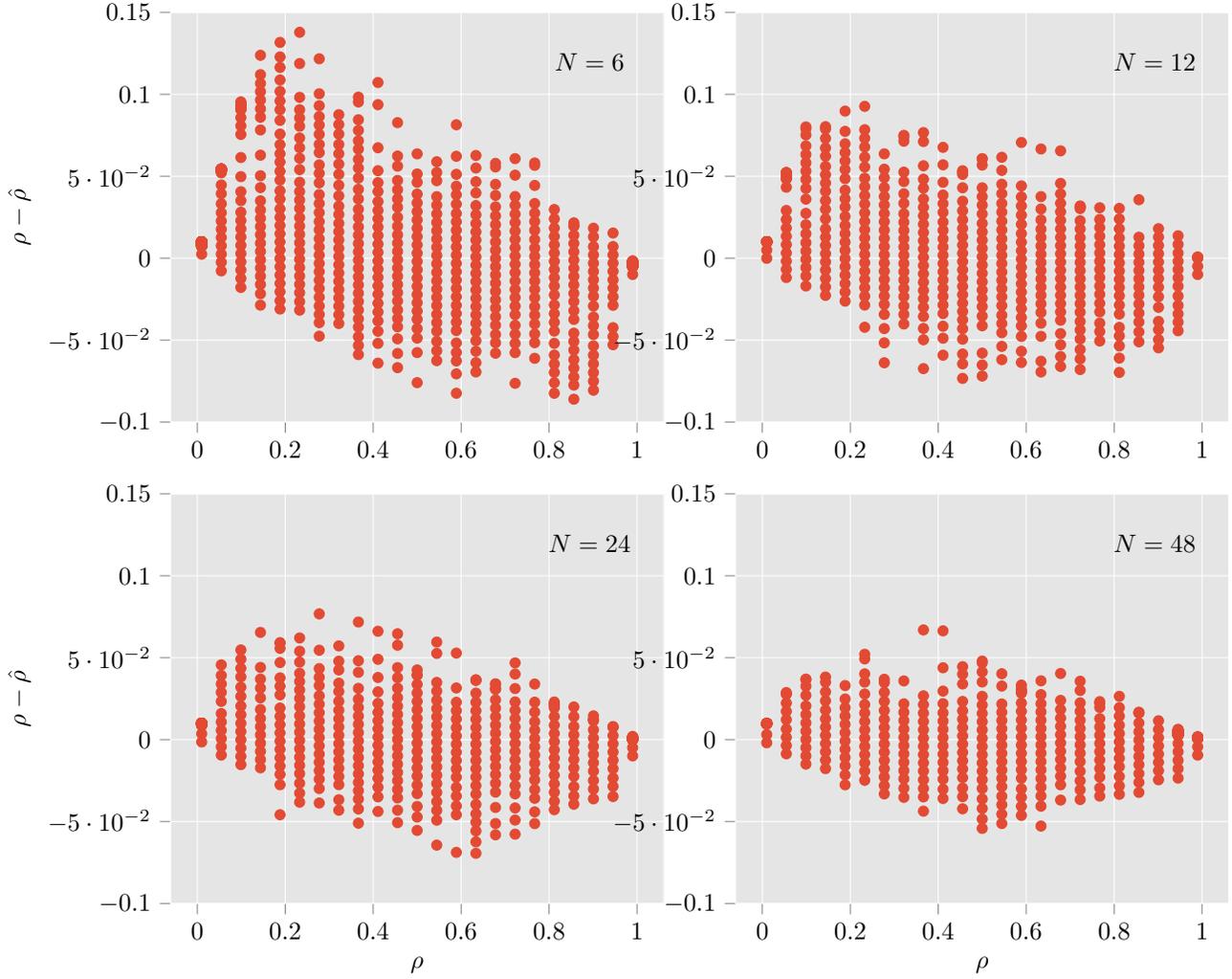}
		\caption{Comparison between $\rho$ and the estimation error calculated as $\rho - \hat{\rho}$ for different number of subjects $N$.} \label{fig:rho_per_n}
	\end{center}
\end{figure*}

Similarly to Fig.~\ref{fig:psi_per_n} for $\psi$ we present Fig.~\ref{fig:rho_per_n} for the $\rho$ parameter. $\rho$ describes variance and therefore the number of subjects influences the obtained results more significantly. For a small number of subjects a strong asymmetry of the results is clearly visible. It is a consequence of the limitations of the parameter and the correct estimation process, which cannot obtain a parameter out if its range. For a large number of subjects it is nearly invisible. Based on the results presented in Fig.~\ref{fig:rho_per_n} the confidence interval for $N=6$ is close to 0.2, which is 20\% of the whole range, for $N=24$ the confidence interval is closer to 0.1. Again the formal analysis of the confidence interval is left as future work.

The implementation of the estimation procedure is provided at  \url{https://colab.research.google.com/drive/1ioM4JqUaEA8fHJH9V-0iphHt4C9zmI0i}. The code allows to estimate parameters from a sample in a general form of $M$ possible categories (see Appendix~\ref{sec:ap:n}). Also a simulation and visualization of the GSD is provided.  

\section{Data Sets}
\label{sec:datasets}

We use four data sets to check practical distribution of subjects' answers. Those data sets come from four subjective tests: (i) VQEG HDTV Phase I \cite{HDTV_Phase_I_test}, (ii) ITS4S \cite{ITS4S_test}, (iii) AGH/NTIA \cite{Janowski2014, AGH_NTIA_14-505} and (iv) MM2 \cite{MM2_test}. This gives a total of 71,212 observations. All the tests utilize the Absolute Category Rating (ACR) method (in accordance with Recommendation ITU-T P.910). Additionally, all subjective scores are given on the following five-level scale: (1) Bad, (2) Poor, (3) Fair, (4) Good and (5) Excellent. All four data sets are publicly accessible on CDVL (www.cdvl.org)\footnote{Search keywords for all four data sets (in the same order as listed in the text): ``vqeghdN'' (replace N with a number from 1 to 6 to find videos from all 6 experiments), ``its4s'', ``AGH/NTIA'' and ``mm2''.}. If available, data from training sessions can be used as well. We do it to make the analysis as general as possible. Following subsections describe each data set.

\subsection{VQEG HDTV Phase I}
\label{ssec:vqeg_hdtv_phase_i}

VQEG HDTV Phase I is a classical data set containing scores from 6 video-only experiments using a controlled viewing environment (in accordance with Recommendation ITU-R BT.500-11) with various display technologies. The display resolution is constant (1920x1080). There are 24 subjects per each experiment, all of whom produce scores for 168 
Processed Video Sequences (PVSs). All experiments use two sets of PVSs: specific and common. The common set is shown in all 
experiments. It contain 24 PVSs. The remaining 144 PVSs are used in the specific set and do not repeat across experiments. 
Those 144 PVSs are created by applying 15 degradation conditions (i.e. Hypothetical Reference Circuits, HRCs) to 9 pristine 
sequences (i.e. Source Reference Sequences, SRCs). All videos are 10 seconds long and contain various coding and transmission 
artifacts. The distortions are meant to simulate a digital transmission of video over a communication channel. The variation of 
the ACR method is used, namely Absolute Category Rating with Hidden Reference (ACR-HR).

\subsection{ITS4S}
\label{ssec:its4s}

Designed for training no-reference (NR) metrics, the ITS4S data set contains 813 unique video sequences (without audio). It 
characterizes a generic adaptive streaming system for high definition mobile devices. All sequences are four seconds long, no 
sequence is repeated in a single experiment, the data set emphasizes original footage (i.e. as created by professional 
producers), 35\% of sequences contain no compression, the remaining 65\% contain only simple compression artifacts. 

All sequences are presented in 720p resolution (1280x720) and frame rate equals to 24. However, some sequences are encoded with lower resolution (e.g. 512x288) and have to be up-scaled to fit the display resolution (i.e. 720p).

The subjective scores included in the data set come from two subjective tests performed by two laboratories in 
two different countries. One laboratory uses subjects hired through a temporary hiring agency. The other collects data using 24 
engineering students.

\subsection{MM2}
\label{ssec:mm2}

An audiovisual test performs in 10 different environments (controlled and public). We treat scores coming 
from different environments as independent. Thus, we assume the data represents 10 independent experiments. Both the 
stimuli (640x480@30fps audiovisual sequences) and scale are held constant between the experiments. The stimuli represents a 
wide range of quality, degradation are introduced solely be compression. Importantly, the encoding levels of audio and video 
streams are chosen to match one of the three encoding qualities: (i) high, (ii) medium and (iii) low. 

Depending on the experiment: 9 to 34 subjects are used, subjects spanned all age groups, some experiments use overlapping 
pools of subjects. Nevertheless, due to the lack of precise subject labeling, we assume all subjects are independent.

\subsection{AGH/NTIA}
\label{ssec:agh-ntia}

AGH/NTIA is a video-only subjective test designed to examine three issues: (i) the behavior of subjects when repeatedly rating the same 
stimuli, (ii) the suitability of subjects screening techniques and (iii) the influence of source sequence reuse on subjects 
scores. Depending on the stimuli, there are 3 to 6 ratings from the same subject. Scores from 29 subjects are available. One of 
those subjects is a video quality expert, who knows the purpose of the experiment. He tries to replicate his previous score for 
any repeating PVS. In addition, two subjects perform the test based on intentionally incorrect instructions. This is done to test subjects' screening methods. 

\section{Subjective Data Analysis}
\label{sec:analysis}

In the previous sections we propose the new distribution and prove that the estimation method extracts data from a simulated sample correctly. In this section we validate if the proposed distribution can be used to model real subjective data. In addition we compare the GSD with the continuous model. 

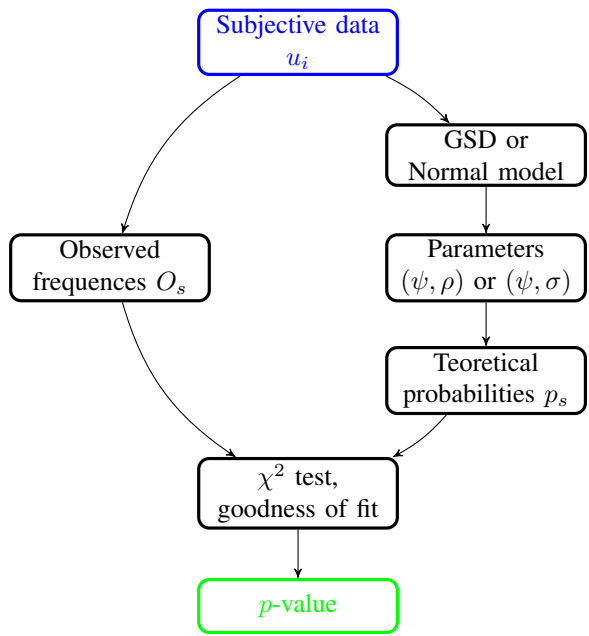
\begin{figure}
	\centering
	\begin{tikzpicture}[->,>=stealth']
	\node (data) [state, color = blue] {\parbox{2.5cm}{\centering Subjective data \\ $u_{i}$}};
	
	\node (dist) [state,    	
	yshift=-1.5cm, 		
	right of=data, 	
	node distance=2.5cm, 	
	anchor=center] {\parbox{2.5cm}{\centering GSD or \\ Normal model}};
	
	\node (par) [state,    	
	below of=dist, 	
	node distance=1.5cm, 	
	anchor=center] {\parbox{2.5cm}{\centering Parameters \\  $(\psi, \rho)$ or $(\psi,\sigma)$}};
	
	\node (prob) [state,   
	below of=par, 	
	node distance=1.5cm, 	
	anchor=center] {\parbox{2.5cm}{\centering Teoretical probabilities $p_s$}};
	
	\node (obs) [state,    	
	yshift=-3cm,
	left of=data, 	
	node distance=2.5cm, 	
	anchor=center] {\parbox{2.5cm}{\centering Observed frequences $O_s$}};
	
	\node (chi2) [state,    	
	yshift=-3cm, 		
	right of=obs, 	
	node distance=2.5cm, 	
	anchor=center] {\parbox{2.5cm}{\centering $\chi^2$ test, \\ goodness of fit}};
	
	\node (pval) [state,    	
	below of=chi2, 	
	node distance=1.5cm, 	
	anchor=center, color = green] {\parbox{2.5cm}{\centering $p$-value}};

	\path 	(data)	edge[bend left=10]  (dist);
	\path (dist) edge (par);
	\path (par) edge (prob);
	\path (data) edge[bend right=20] (obs);
	\path 	(prob)	edge[bend left=10]  (chi2);
	\path (obs) edge[bend right=20] (chi2);
	\path (chi2) edge (pval);
	\end{tikzpicture}
	
	\caption{Algorithm used to test goodness-of-fit. The pipeline starts at the blue ``Subjective data`` block. Its output is a $p$-value (green box), which we use to verify the null hypothesis, that a sample comes from the assumed distribution (either GSD or the normal model)}\label{fig:test}
\end{figure}

In order to validate if the distribution fits specific data we have to perform a two-steps procedure. The first step is to estimate the distribution parameters from a sample. The second step is to test a null hypothesis that the sample comes from estimated distribution. Fig.~\ref{fig:test} visualizes the procedure. Since we have discrete data the most natural testing tool is Pearson's $\chi^2$ test, also called goodness-of-fit test (see \cite{Walpole_Probablity_2007}). 

\begin{figure}
	\begin{center}
		\input{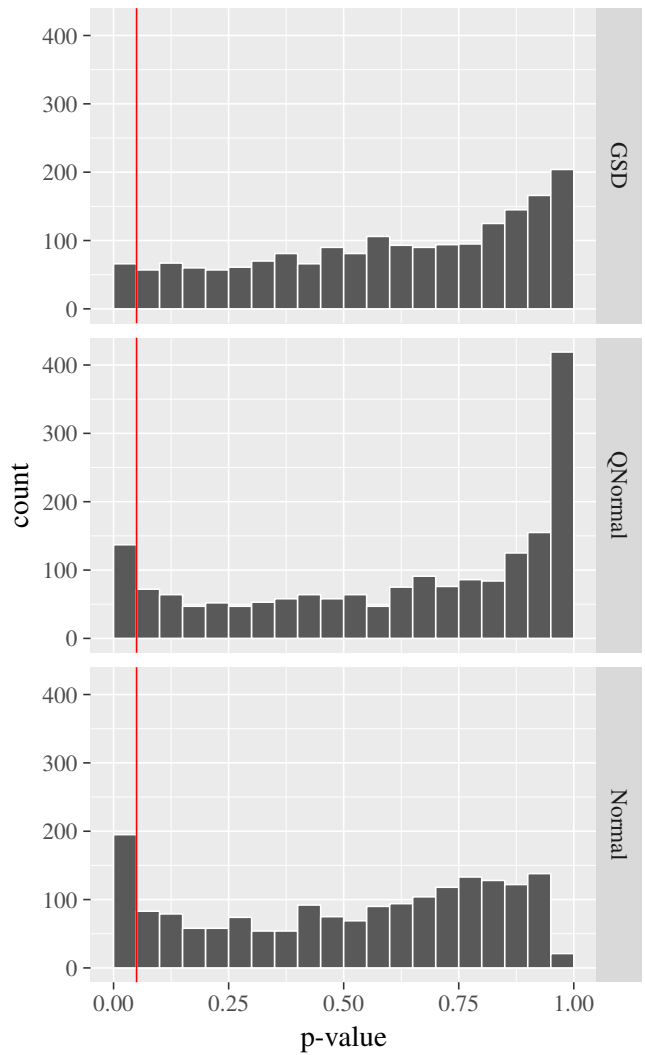}
		\caption{$p$-value obtained for different distributions. The red line marks the 0.05 significance level.} \label{fig:pval}
	\end{center}
\end{figure}

In Section~\ref{sec:datasets} we describe four different databases. Here we are interested in a single PVS (video sequence) analysis, we are not taking into account from which database specific PVS comes from. The key parameter is a number of PVSs. There are 1,879 PVSs with at least 24 answers each (some PVSs have 213 answers). We have five PVSs with all 24 answers being one. We remove them from the further analysis since for the continuous model we obtain NaNs. This results in 1,874 PVSs available for the analysis. For each PVS we estimate the GSD (or continuous model) parameters and, finally, calculated $p$-value of the goodness-of-fit test. We show results for three cases: (i) the GSD, (ii) the quantized continuous model (labelled \textit{QNormal}) and (iii) the continuous model with its parameters estimated by sample mean and standard deviation (labelled \textit{Normal}). For each case we obtain 1,874 $p$-values. Fig.~\ref{fig:pval} compares $p$-values distributions for all three cases.

We use $p$-values (from multiple hypothesis tests) to test a general hypothesis. The procedure is as follows: for a single PVS we obtain $p$-value $a$, if $a > \alpha$ one can assume that this PVS distribution comes from a distribution, for example GSD. This result is true for this particular PVS, but we are interested in a general PVS distribution. Thus, we have to run this test for a large number of PVSs, obtaining a large number of $p$-values $a_i$. If we repeat this hypothesis testing we know that some of the tests fails by chance. The exact number depends on the test difficulty. If a test is ``difficult'' it fails with a higher probability than an ``easier'' test. Hypothesis testing is constructed in such a way that even for ``difficult'' cases failure probability does not exceed $\alpha$. Therefore, for PVSs following a specific distribution (i.e., the one assumed in the null hypothesis) we should have: 
\begin{equation} \label{eq:p}
P(A_i < \alpha) \leq \alpha,
\end{equation}
where $A_i$ is a random variable describing the obtained $p$-values and $\alpha$ is a significance level. 

This can be satisfied by various $p$-values distributions. Obviously, if all $p$-values are close to 1, equation (\ref{eq:p}) is true. If $A_i$ has uniform distribution, (\ref{eq:p}) is also true. Nevertheless, for any increase, beyond what is allowed by the uniform distribution, in the probability for small $p$-values, results in rejecting the global hypothesis about the specific distribution.

The $p$-values histogram for the GSD in Fig.~\ref{fig:pval} shows an increasing probability of a given $p$-value as a function of $p$-value. As described above, it shows a strong evidence that the global hypothesis is true. The probability of observing $p$-value smaller than 0.05 is 0.035, which is in line with the theory. However, for both continuous models we obtain $p$-value smaller than 0.05 with probability 0.073 for QNormal and 0.106 for the Normal model. Moreover, for Normal and QNormal in Fig.~\ref{fig:pval} we can see that probability for the smallest $p$-value is clearly higher than for slightly larger $p$-values. In both cases the observed answers are not coming from those distributions.

We run the following test to formally prove the global hypothesis
$$H_0: P(A<\alpha) \leq \alpha$$
versus 
$$H_1: P(A<\alpha)>\alpha,$$ 
which can be interpreted as a test of $H_0$: the model is correct versus $H_1$: the model is incorrect. If we denote by $p_{\text{GSD}}$, $p_{\text{QNormal}}$ and $p_{\text{Normal}}$ the $p$-values of the above test for GSD, QNormal and Normal models respectively, we obtain for $\alpha=0.05$:
$$p_{\text{GSD}}=0.997252727,$$
$$p_{\text{QNormal}}=0.000005612,$$
$$p_{\text{Normal}}=0.000000000$$
It clearly shows that for QNormal and Normal models too many $p$-values are below $0.05$ to say those models are probable.  

The above result is very important from the further usage of the GSD in the context of video quality subjective testing perspective. We demonstrate that subjective answers from different subjective tests are following the GSD distribution. It means that GSD is the correct way of modeling subjective scores in video quality experiments. We strongly believe that GSD can be used in different cases of experiments with answers on a Likert scale.  

\subsection{A Priori Distributions}

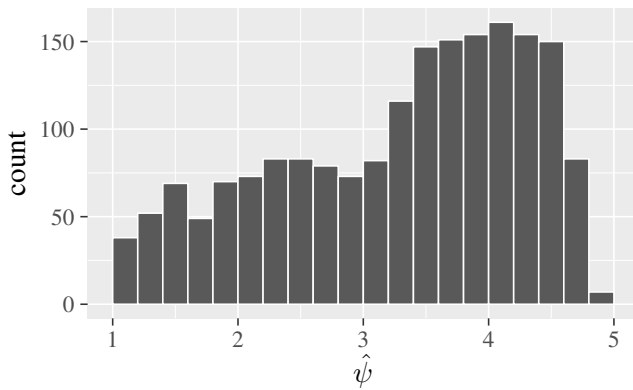
\begin{figure}
	\begin{center}
\begin{tikzpicture}[x=1pt,y=1pt]
\definecolor{fillColor}{RGB}{255,255,255}
\path[use as bounding box,fill=fillColor,fill opacity=0.00] (0,0) rectangle (250.78,154.99);
\begin{scope}
\path[clip] (  0.00,  0.00) rectangle (250.78,154.99);
\definecolor{drawColor}{RGB}{255,255,255}
\definecolor{fillColor}{RGB}{255,255,255}

\path[draw=drawColor,line width= 0.6pt,line join=round,line cap=round,fill=fillColor] (  0.00,  0.00) rectangle (250.78,154.99);
\end{scope}
\begin{scope}
\path[clip] ( 36.72, 32.09) rectangle (245.28,149.49);
\definecolor{fillColor}{gray}{0.92}

\path[fill=fillColor] ( 36.72, 32.09) rectangle (245.28,149.49);
\definecolor{drawColor}{RGB}{255,255,255}

\path[draw=drawColor,line width= 0.3pt,line join=round] ( 36.72, 54.00) --
	(245.28, 54.00);

\path[draw=drawColor,line width= 0.3pt,line join=round] ( 36.72, 87.14) --
	(245.28, 87.14);

\path[draw=drawColor,line width= 0.3pt,line join=round] ( 36.72,120.29) --
	(245.28,120.29);

\path[draw=drawColor,line width= 0.3pt,line join=round] ( 69.90, 32.09) --
	( 69.90,149.49);

\path[draw=drawColor,line width= 0.3pt,line join=round] (117.30, 32.09) --
	(117.30,149.49);

\path[draw=drawColor,line width= 0.3pt,line join=round] (164.70, 32.09) --
	(164.70,149.49);

\path[draw=drawColor,line width= 0.3pt,line join=round] (212.10, 32.09) --
	(212.10,149.49);

\path[draw=drawColor,line width= 0.6pt,line join=round] ( 36.72, 37.42) --
	(245.28, 37.42);

\path[draw=drawColor,line width= 0.6pt,line join=round] ( 36.72, 70.57) --
	(245.28, 70.57);

\path[draw=drawColor,line width= 0.6pt,line join=round] ( 36.72,103.72) --
	(245.28,103.72);

\path[draw=drawColor,line width= 0.6pt,line join=round] ( 36.72,136.86) --
	(245.28,136.86);

\path[draw=drawColor,line width= 0.6pt,line join=round] ( 46.20, 32.09) --
	( 46.20,149.49);

\path[draw=drawColor,line width= 0.6pt,line join=round] ( 93.60, 32.09) --
	( 93.60,149.49);

\path[draw=drawColor,line width= 0.6pt,line join=round] (141.00, 32.09) --
	(141.00,149.49);

\path[draw=drawColor,line width= 0.6pt,line join=round] (188.40, 32.09) --
	(188.40,149.49);

\path[draw=drawColor,line width= 0.6pt,line join=round] (235.80, 32.09) --
	(235.80,149.49);
\definecolor{fillColor}{gray}{0.35}

\path[draw=drawColor,line width= 0.6pt,line join=round,fill=fillColor] ( 46.20, 37.42) rectangle ( 55.68, 62.61);

\path[draw=drawColor,line width= 0.6pt,line join=round,fill=fillColor] ( 55.68, 37.42) rectangle ( 65.16, 71.90);

\path[draw=drawColor,line width= 0.6pt,line join=round,fill=fillColor] ( 65.16, 37.42) rectangle ( 74.64, 83.17);

\path[draw=drawColor,line width= 0.6pt,line join=round,fill=fillColor] ( 74.64, 37.42) rectangle ( 84.12, 69.91);

\path[draw=drawColor,line width= 0.6pt,line join=round,fill=fillColor] ( 84.12, 37.42) rectangle ( 93.60, 83.83);

\path[draw=drawColor,line width= 0.6pt,line join=round,fill=fillColor] ( 93.60, 37.42) rectangle (103.08, 85.82);

\path[draw=drawColor,line width= 0.6pt,line join=round,fill=fillColor] (103.08, 37.42) rectangle (112.56, 92.45);

\path[draw=drawColor,line width= 0.6pt,line join=round,fill=fillColor] (112.56, 37.42) rectangle (122.04, 92.45);

\path[draw=drawColor,line width= 0.6pt,line join=round,fill=fillColor] (122.04, 37.42) rectangle (131.52, 89.79);

\path[draw=drawColor,line width= 0.6pt,line join=round,fill=fillColor] (131.52, 37.42) rectangle (141.00, 85.82);

\path[draw=drawColor,line width= 0.6pt,line join=round,fill=fillColor] (141.00, 37.42) rectangle (150.48, 91.78);

\path[draw=drawColor,line width= 0.6pt,line join=round,fill=fillColor] (150.48, 37.42) rectangle (159.96,114.32);

\path[draw=drawColor,line width= 0.6pt,line join=round,fill=fillColor] (159.96, 37.42) rectangle (169.44,134.87);

\path[draw=drawColor,line width= 0.6pt,line join=round,fill=fillColor] (169.44, 37.42) rectangle (178.92,137.53);

\path[draw=drawColor,line width= 0.6pt,line join=round,fill=fillColor] (178.92, 37.42) rectangle (188.40,139.51);

\path[draw=drawColor,line width= 0.6pt,line join=round,fill=fillColor] (188.40, 37.42) rectangle (197.88,144.16);

\path[draw=drawColor,line width= 0.6pt,line join=round,fill=fillColor] (197.88, 37.42) rectangle (207.36,139.51);

\path[draw=drawColor,line width= 0.6pt,line join=round,fill=fillColor] (207.36, 37.42) rectangle (216.84,136.86);

\path[draw=drawColor,line width= 0.6pt,line join=round,fill=fillColor] (216.84, 37.42) rectangle (226.32, 92.45);

\path[draw=drawColor,line width= 0.6pt,line join=round,fill=fillColor] (226.32, 37.42) rectangle (235.80, 42.06);
\end{scope}
\begin{scope}
\path[clip] (  0.00,  0.00) rectangle (250.78,154.99);
\definecolor{drawColor}{gray}{0.30}

\node[text=drawColor,anchor=base east,inner sep=0pt, outer sep=0pt, scale=  0.88] at ( 31.77, 34.39) {0};

\node[text=drawColor,anchor=base east,inner sep=0pt, outer sep=0pt, scale=  0.88] at ( 31.77, 67.54) {50};

\node[text=drawColor,anchor=base east,inner sep=0pt, outer sep=0pt, scale=  0.88] at ( 31.77,100.69) {100};

\node[text=drawColor,anchor=base east,inner sep=0pt, outer sep=0pt, scale=  0.88] at ( 31.77,133.83) {150};
\end{scope}
\begin{scope}
\path[clip] (  0.00,  0.00) rectangle (250.78,154.99);
\definecolor{drawColor}{gray}{0.20}

\path[draw=drawColor,line width= 0.6pt,line join=round] ( 33.97, 37.42) --
	( 36.72, 37.42);

\path[draw=drawColor,line width= 0.6pt,line join=round] ( 33.97, 70.57) --
	( 36.72, 70.57);

\path[draw=drawColor,line width= 0.6pt,line join=round] ( 33.97,103.72) --
	( 36.72,103.72);

\path[draw=drawColor,line width= 0.6pt,line join=round] ( 33.97,136.86) --
	( 36.72,136.86);
\end{scope}
\begin{scope}
\path[clip] (  0.00,  0.00) rectangle (250.78,154.99);
\definecolor{drawColor}{gray}{0.20}

\path[draw=drawColor,line width= 0.6pt,line join=round] ( 46.20, 29.34) --
	( 46.20, 32.09);

\path[draw=drawColor,line width= 0.6pt,line join=round] ( 93.60, 29.34) --
	( 93.60, 32.09);

\path[draw=drawColor,line width= 0.6pt,line join=round] (141.00, 29.34) --
	(141.00, 32.09);

\path[draw=drawColor,line width= 0.6pt,line join=round] (188.40, 29.34) --
	(188.40, 32.09);

\path[draw=drawColor,line width= 0.6pt,line join=round] (235.80, 29.34) --
	(235.80, 32.09);
\end{scope}
\begin{scope}
\path[clip] (  0.00,  0.00) rectangle (250.78,154.99);
\definecolor{drawColor}{gray}{0.30}

\node[text=drawColor,anchor=base,inner sep=0pt, outer sep=0pt, scale=  0.88] at ( 46.20, 21.08) {1};

\node[text=drawColor,anchor=base,inner sep=0pt, outer sep=0pt, scale=  0.88] at ( 93.60, 21.08) {2};

\node[text=drawColor,anchor=base,inner sep=0pt, outer sep=0pt, scale=  0.88] at (141.00, 21.08) {3};

\node[text=drawColor,anchor=base,inner sep=0pt, outer sep=0pt, scale=  0.88] at (188.40, 21.08) {4};

\node[text=drawColor,anchor=base,inner sep=0pt, outer sep=0pt, scale=  0.88] at (235.80, 21.08) {5};
\end{scope}
\begin{scope}
\path[clip] (  0.00,  0.00) rectangle (250.78,154.99);
\definecolor{drawColor}{RGB}{0,0,0}

\node[text=drawColor,anchor=base,inner sep=0pt, outer sep=0pt, scale=  1.10] at (141.00,  8.00) {$\hat\psi$};
\end{scope}
\begin{scope}
\path[clip] (  0.00,  0.00) rectangle (250.78,154.99);
\definecolor{drawColor}{RGB}{0,0,0}

\node[text=drawColor,rotate= 90.00,anchor=base,inner sep=0pt, outer sep=0pt, scale=  1.10] at ( 13.08, 90.79) {count};
\end{scope}
\end{tikzpicture}
		\caption{Distribution of $\hat{\psi}$ for all 1,874 PVSs.} \label{fig:psihathist}
	\end{center}
\end{figure}

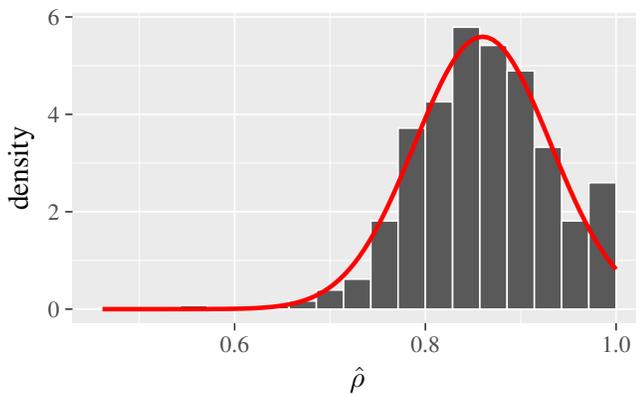
\begin{figure}
	\begin{center}
\begin{tikzpicture}[x=1pt,y=1pt]
\definecolor{fillColor}{RGB}{255,255,255}
\path[use as bounding box,fill=fillColor,fill opacity=0.00] (0,0) rectangle (250.78,154.99);
\begin{scope}
\path[clip] (  0.00,  0.00) rectangle (250.78,154.99);
\definecolor{drawColor}{RGB}{255,255,255}
\definecolor{fillColor}{RGB}{255,255,255}

\path[draw=drawColor,line width= 0.6pt,line join=round,line cap=round,fill=fillColor] (  0.00,  0.00) rectangle (250.78,154.99);
\end{scope}
\begin{scope}
\path[clip] ( 29.87, 32.09) rectangle (245.28,149.49);
\definecolor{fillColor}{gray}{0.92}

\path[fill=fillColor] ( 29.87, 32.09) rectangle (245.28,149.49);
\definecolor{drawColor}{RGB}{255,255,255}

\path[draw=drawColor,line width= 0.3pt,line join=round] ( 29.87, 55.86) --
	(245.28, 55.86);

\path[draw=drawColor,line width= 0.3pt,line join=round] ( 29.87, 92.73) --
	(245.28, 92.73);

\path[draw=drawColor,line width= 0.3pt,line join=round] ( 29.87,129.60) --
	(245.28,129.60);

\path[draw=drawColor,line width= 0.3pt,line join=round] ( 55.12, 32.09) --
	( 55.12,149.49);

\path[draw=drawColor,line width= 0.3pt,line join=round] (127.27, 32.09) --
	(127.27,149.49);

\path[draw=drawColor,line width= 0.3pt,line join=round] (199.41, 32.09) --
	(199.41,149.49);

\path[draw=drawColor,line width= 0.6pt,line join=round] ( 29.87, 37.42) --
	(245.28, 37.42);

\path[draw=drawColor,line width= 0.6pt,line join=round] ( 29.87, 74.29) --
	(245.28, 74.29);

\path[draw=drawColor,line width= 0.6pt,line join=round] ( 29.87,111.16) --
	(245.28,111.16);

\path[draw=drawColor,line width= 0.6pt,line join=round] ( 29.87,148.03) --
	(245.28,148.03);

\path[draw=drawColor,line width= 0.6pt,line join=round] ( 91.19, 32.09) --
	( 91.19,149.49);

\path[draw=drawColor,line width= 0.6pt,line join=round] (163.34, 32.09) --
	(163.34,149.49);

\path[draw=drawColor,line width= 0.6pt,line join=round] (235.49, 32.09) --
	(235.49,149.49);
\definecolor{fillColor}{gray}{0.35}

\path[draw=drawColor,line width= 0.6pt,line join=round,fill=fillColor] ( 39.66, 37.42) rectangle ( 49.97, 37.77);

\path[draw=drawColor,line width= 0.6pt,line join=round,fill=fillColor] ( 49.97, 37.42) rectangle ( 60.27, 37.42);

\path[draw=drawColor,line width= 0.6pt,line join=round,fill=fillColor] ( 60.27, 37.42) rectangle ( 70.58, 37.42);

\path[draw=drawColor,line width= 0.6pt,line join=round,fill=fillColor] ( 70.58, 37.42) rectangle ( 80.89, 38.80);

\path[draw=drawColor,line width= 0.6pt,line join=round,fill=fillColor] ( 80.89, 37.42) rectangle ( 91.19, 37.42);

\path[draw=drawColor,line width= 0.6pt,line join=round,fill=fillColor] ( 91.19, 37.42) rectangle (101.50, 38.46);

\path[draw=drawColor,line width= 0.6pt,line join=round,fill=fillColor] (101.50, 37.42) rectangle (111.81, 38.46);

\path[draw=drawColor,line width= 0.6pt,line join=round,fill=fillColor] (111.81, 37.42) rectangle (122.11, 40.52);

\path[draw=drawColor,line width= 0.6pt,line join=round,fill=fillColor] (122.11, 37.42) rectangle (132.42, 44.65);

\path[draw=drawColor,line width= 0.6pt,line join=round,fill=fillColor] (132.42, 37.42) rectangle (142.73, 48.78);

\path[draw=drawColor,line width= 0.6pt,line join=round,fill=fillColor] (142.73, 37.42) rectangle (153.03, 70.82);

\path[draw=drawColor,line width= 0.6pt,line join=round,fill=fillColor] (153.03, 37.42) rectangle (163.34,105.94);

\path[draw=drawColor,line width= 0.6pt,line join=round,fill=fillColor] (163.34, 37.42) rectangle (173.65,115.92);

\path[draw=drawColor,line width= 0.6pt,line join=round,fill=fillColor] (173.65, 37.42) rectangle (183.95,144.16);

\path[draw=drawColor,line width= 0.6pt,line join=round,fill=fillColor] (183.95, 37.42) rectangle (194.26,137.27);

\path[draw=drawColor,line width= 0.6pt,line join=round,fill=fillColor] (194.26, 37.42) rectangle (204.57,127.63);

\path[draw=drawColor,line width= 0.6pt,line join=round,fill=fillColor] (204.57, 37.42) rectangle (214.87, 98.71);

\path[draw=drawColor,line width= 0.6pt,line join=round,fill=fillColor] (214.87, 37.42) rectangle (225.18, 70.82);

\path[draw=drawColor,line width= 0.6pt,line join=round,fill=fillColor] (225.18, 37.42) rectangle (235.49, 85.28);
\definecolor{drawColor}{RGB}{255,0,0}

\path[draw=drawColor,line width= 1.7pt,line join=round] ( 41.21, 37.42) --
	( 43.16, 37.42) --
	( 45.10, 37.42) --
	( 47.04, 37.42) --
	( 48.99, 37.42) --
	( 50.93, 37.42) --
	( 52.87, 37.42) --
	( 54.81, 37.42) --
	( 56.76, 37.42) --
	( 58.70, 37.42) --
	( 60.64, 37.42) --
	( 62.58, 37.42) --
	( 64.53, 37.42) --
	( 66.47, 37.43) --
	( 68.41, 37.43) --
	( 70.36, 37.43) --
	( 72.30, 37.43) --
	( 74.24, 37.43) --
	( 76.18, 37.44) --
	( 78.13, 37.44) --
	( 80.07, 37.45) --
	( 82.01, 37.46) --
	( 83.95, 37.47) --
	( 85.90, 37.48) --
	( 87.84, 37.50) --
	( 89.78, 37.53) --
	( 91.73, 37.57) --
	( 93.67, 37.61) --
	( 95.61, 37.67) --
	( 97.55, 37.74) --
	( 99.50, 37.83) --
	(101.44, 37.95) --
	(103.38, 38.09) --
	(105.32, 38.27) --
	(107.27, 38.49) --
	(109.21, 38.76) --
	(111.15, 39.09) --
	(113.09, 39.49) --
	(115.04, 39.97) --
	(116.98, 40.54) --
	(118.92, 41.22) --
	(120.87, 42.02) --
	(122.81, 42.96) --
	(124.75, 44.05) --
	(126.69, 45.31) --
	(128.64, 46.76) --
	(130.58, 48.41) --
	(132.52, 50.27) --
	(134.46, 52.37) --
	(136.41, 54.71) --
	(138.35, 57.31) --
	(140.29, 60.16) --
	(142.24, 63.28) --
	(144.18, 66.66) --
	(146.12, 70.29) --
	(148.06, 74.16) --
	(150.01, 78.25) --
	(151.95, 82.54) --
	(153.89, 87.00) --
	(155.83, 91.59) --
	(157.78, 96.27) --
	(159.72,100.99) --
	(161.66,105.70) --
	(163.61,110.33) --
	(165.55,114.85) --
	(167.49,119.17) --
	(169.43,123.24) --
	(171.38,127.00) --
	(173.32,130.40) --
	(175.26,133.38) --
	(177.20,135.89) --
	(179.15,137.89) --
	(181.09,139.35) --
	(183.03,140.24) --
	(184.98,140.55) --
	(186.92,140.27) --
	(188.86,139.41) --
	(190.80,137.98) --
	(192.75,136.01) --
	(194.69,133.52) --
	(196.63,130.57) --
	(198.57,127.19) --
	(200.52,123.45) --
	(202.46,119.39) --
	(204.40,115.08) --
	(206.34,110.58) --
	(208.29,105.95) --
	(210.23,101.24) --
	(212.17, 96.52) --
	(214.12, 91.84) --
	(216.06, 87.24) --
	(218.00, 82.78) --
	(219.94, 78.47) --
	(221.89, 74.37) --
	(223.83, 70.49) --
	(225.77, 66.84) --
	(227.71, 63.45) --
	(229.66, 60.32) --
	(231.60, 57.45) --
	(233.54, 54.84) --
	(235.49, 52.49);
\end{scope}
\begin{scope}
\path[clip] (  0.00,  0.00) rectangle (250.78,154.99);
\definecolor{drawColor}{gray}{0.30}

\node[text=drawColor,anchor=base east,inner sep=0pt, outer sep=0pt, scale=  0.88] at ( 24.92, 34.39) {0};

\node[text=drawColor,anchor=base east,inner sep=0pt, outer sep=0pt, scale=  0.88] at ( 24.92, 71.26) {2};

\node[text=drawColor,anchor=base east,inner sep=0pt, outer sep=0pt, scale=  0.88] at ( 24.92,108.13) {4};

\node[text=drawColor,anchor=base east,inner sep=0pt, outer sep=0pt, scale=  0.88] at ( 24.92,145.00) {6};
\end{scope}
\begin{scope}
\path[clip] (  0.00,  0.00) rectangle (250.78,154.99);
\definecolor{drawColor}{gray}{0.20}

\path[draw=drawColor,line width= 0.6pt,line join=round] ( 27.12, 37.42) --
	( 29.87, 37.42);

\path[draw=drawColor,line width= 0.6pt,line join=round] ( 27.12, 74.29) --
	( 29.87, 74.29);

\path[draw=drawColor,line width= 0.6pt,line join=round] ( 27.12,111.16) --
	( 29.87,111.16);

\path[draw=drawColor,line width= 0.6pt,line join=round] ( 27.12,148.03) --
	( 29.87,148.03);
\end{scope}
\begin{scope}
\path[clip] (  0.00,  0.00) rectangle (250.78,154.99);
\definecolor{drawColor}{gray}{0.20}

\path[draw=drawColor,line width= 0.6pt,line join=round] ( 91.19, 29.34) --
	( 91.19, 32.09);

\path[draw=drawColor,line width= 0.6pt,line join=round] (163.34, 29.34) --
	(163.34, 32.09);

\path[draw=drawColor,line width= 0.6pt,line join=round] (235.49, 29.34) --
	(235.49, 32.09);
\end{scope}
\begin{scope}
\path[clip] (  0.00,  0.00) rectangle (250.78,154.99);
\definecolor{drawColor}{gray}{0.30}

\node[text=drawColor,anchor=base,inner sep=0pt, outer sep=0pt, scale=  0.88] at ( 91.19, 21.08) {0.6};

\node[text=drawColor,anchor=base,inner sep=0pt, outer sep=0pt, scale=  0.88] at (163.34, 21.08) {0.8};

\node[text=drawColor,anchor=base,inner sep=0pt, outer sep=0pt, scale=  0.88] at (235.49, 21.08) {1.0};
\end{scope}
\begin{scope}
\path[clip] (  0.00,  0.00) rectangle (250.78,154.99);
\definecolor{drawColor}{RGB}{0,0,0}

\node[text=drawColor,anchor=base,inner sep=0pt, outer sep=0pt, scale=  1.10] at (137.57,  8.00) {$\hat\rho$};
\end{scope}
\begin{scope}
\path[clip] (  0.00,  0.00) rectangle (250.78,154.99);
\definecolor{drawColor}{RGB}{0,0,0}

\node[text=drawColor,rotate= 90.00,anchor=base,inner sep=0pt, outer sep=0pt, scale=  1.10] at ( 13.08, 90.79) {density};
\end{scope}
\end{tikzpicture}
		\caption{Distribution of $\hat{\rho}$ for all 1,874 PVSs. Red line shows the normal distribution fitted to the obtained data.} \label{fig:rhohatdist}
	\end{center}
\end{figure}

Since we have an evidence that the GSD is the correct way of modeling subjective scores it is important for further analysis to describe the typical parameters of this distribution. We have two parameters: $\psi$ (see Fig.~\ref{fig:psihathist}) and $\rho$ (see. Fig.~\ref{fig:rhohatdist}). 

The first parameter depends on the experiment setup. In a typical situation its distribution (defined by many PVSs utilized in the experiment) should be close to uniform distribution. What we can see is a lack of very high qualities (close to 5). Another observation is a greater number of high quality PVSs than lower quality ones. We theorize it is just an artifact of the type of database we use. 

Much more interesting is the $\rho$ distribution. This parameter is limited to (0,1] interval. We can see that Normal distribution approximates it with a high accuracy. We do not perform a formal test since we know it would be rejected by the high probability of ones. The analysis of QQ plot shows that normal distribution with values higher than one truncated to one, with parameters $\mu_{\rho}=0.86$ and $\sigma_{\rho}=0.071$, can be used to simulate subjective scores. 

\section{Conclusion}

In this paper we propose Generalized Score Distribution (GSD), which is a general form of a discrete distribution with: finite support, two parameters, and not more than one change in the probability monotonicity. The distribution parameters are: $\psi$ determining the mean, and $\rho$ determining the spread of the answers. We focus on the subjective experiments for videos, where subjects score the quality of video sequences on a five point scale. We analyze 1,874 video sequences from publicly available databases, and show with probability of 99.7\%, that subjective answers from those databases are drawn from the GSD distribution. This result opens a new way of analyzing and simulating subjective scores. 

The advantage of GSD is that $\rho$ parameter can be used to determine the type of the underlining process. With $\rho$ close to 1 we know that the process is similar to Bernoulli distribution, for $\rho > C(\psi)$ we know it is Beta Binomial distribution. This information can be used as a diagnostic information: what is the spread of the answers. Note that GSD can be easily used outside of the video quality domain where information about the answer spread is important.  

In the future research we would like to generalize the GSD distribution by adding more parameters similar to \cite{Janowski2015, Li2017}. Another interesting direction of research would be using GSD for data other than subjective experiments for video. We would like collaborate with scientist working in different fields: from audio and image quality to psychology and sociology. In all those fields, a proper modeling of answering process would help to gain new insights. Our results obtained for video quality subjective testing might be used as a proof of concept for those other domains.

\section*{Acknowledgment}

The authors would like to thank Netflix, Inc. for sponsoring this research. This work was supported by the Polish Ministry of Science and Higher Education with the subvention funds of the Faculty of Computer Science, Electronics and Telecommunications of AGH University.

\bibliographystyle{IEEEtran}

\bibliography{main}

\begin{thebibliography}{10}
\providecommand{\url}[1]{#1}
\csname url@samestyle\endcsname
\providecommand{\newblock}{\relax}
\providecommand{\bibinfo}[2]{#2}
\providecommand{\BIBentrySTDinterwordspacing}{\spaceskip=0pt\relax}
\providecommand{\BIBentryALTinterwordstretchfactor}{4}
\providecommand{\BIBentryALTinterwordspacing}{\spaceskip=\fontdimen2\font plus
\BIBentryALTinterwordstretchfactor\fontdimen3\font minus
  \fontdimen4\font\relax}
\providecommand{\BIBforeignlanguage}[2]{{%
\expandafter\ifx\csname l@#1\endcsname\relax
\typeout{** WARNING: IEEEtran.bst: No hyphenation pattern has been}%
\typeout{** loaded for the language `#1'. Using the pattern for}%
\typeout{** the default language instead.}%
\else
\language=\csname l@#1\endcsname
\fi
#2}}
\providecommand{\BIBdecl}{\relax}
\BIBdecl

\bibitem{Tobias_no_silver_bullet_2017}
T.~{Hoßfeld}, P.~E. {Heegaard}, L.~{Skorin-Kapov}, and M.~{Varela}, ``No
  silver bullet: Qoe metrics, qoe fairness, and user diversity in the context
  of qoe management,'' in \emph{2017 Ninth International Conference on Quality
  of Multimedia Experience (QoMEX)}, May 2017, pp. 1--6.

\bibitem{ITUP1401}
\BIBentryALTinterwordspacing
{International Telecommunication Union}, ``{ITU-T Recommendation P.1401:
  Methods, metrics and procedures for statistical evaluation, qualification and
  comparison of objective quality prediction models},'' Tech. Rep., 2012.
  [Online]. Available: \url{https://www.itu.int/rec/T-REC-P.1401-201207-I/en}
\BIBentrySTDinterwordspacing

\bibitem{Brunnstrom2018}
\BIBentryALTinterwordspacing
K.~Brunnstr{\"{o}}m and M.~Barkowsky, ``{Statistical quality of experience
  analysis: on planning the sample size and statistical significance
  testing},'' \emph{Journal of Electronic Imaging}, vol.~27, no.~05, p.~1,
  2018. [Online]. Available:
  \url{https://www.spiedigitallibrary.org/journals/journal-of-electronic-imaging/volume-27/issue-05/053013/Statistical-quality-of-experience-analysis--on-planning-the-sample/10.1117/1.JEI.27.5.053013.full}
\BIBentrySTDinterwordspacing

\bibitem{Hossfeld2016}
\BIBentryALTinterwordspacing
T.~Hossfeld, P.~E. Heegaard, M.~Varela, and S.~M{\"{o}}ller, ``{Formal
  Definition of QoE Metrics},'' pp. 1--23, 2016. [Online]. Available:
  \url{http://arxiv.org/abs/1607.00321{\%}0Ahttp://dx.doi.org/10.1007/s41233-016-0002-1}
\BIBentrySTDinterwordspacing

\bibitem{Janowski2009}
L.~Janowski and Z.~Papir, ``{Modeling subjective tests of quality of experience
  with a generalized linear model},'' \emph{2009 International Workshop on
  Quality of Multimedia Experience, QoMEx 2009}, pp. 35--40, 2009.

\bibitem{Seufert_Fundamental_2019}
M.~{Seufert}, ``Fundamental advantages of considering quality of experience
  distributions over mean opinion scores,'' in \emph{2019 Eleventh
  International Conference on Quality of Multimedia Experience (QoMEX)}, June
  2019, pp. 1--6.

\bibitem{Fiedler2010}
M.~Fiedler, T.~Hossfeld, and P.~Tran-Gia, ``{A generic quantitative
  relationship between quality of experience and quality of service},''
  \emph{IEEE Network}, vol.~24, no.~2, pp. 36--41, 2010.

\bibitem{Hossfeld2011}
T.~Ho{\ss}feld, R.~Schatz, and S.~Egger, ``{SOS: The MOS is not enough!}''
  \emph{2011 3rd International Workshop on Quality of Multimedia Experience,
  QoMEX 2011}, pp. 131--136, 2011.

\bibitem{Hossfeld2018}
\BIBentryALTinterwordspacing
T.~Hossfeld, P.~E. Heegaard, M.~Varela, and L.~Skorin-Kapov, ``{Confidence
  Interval Estimators for MOS Values},'' 2018. [Online]. Available:
  \url{http://arxiv.org/abs/1806.01126}
\BIBentrySTDinterwordspacing

\bibitem{Janowski2015}
L.~Janowski and M.~Pinson, ``The accuracy of subjects in a quality experiment:
  A theoretical subject model,'' \emph{IEEE Transactions on Multimedia},
  vol.~17, no.~12, pp. 2210--2224, Dec 2015.

\bibitem{Li2017}
Z.~Li and C.~G. Bampis, ``{Recover Subjective Quality Scores from Noisy
  Measurements},'' \emph{Data Compression Conference Proceedings}, vol. Part
  F127767, pp. 52--61, 2017.

\bibitem{Kumcu_2017_Performace_of_four}
A.~{Kumcu}, K.~{Bombeke}, L.~{Platiša}, L.~{Jovanov}, J.~{Van Looy}, and
  W.~{Philips}, ``Performance of four subjective video quality assessment
  protocols and impact of different rating preprocessing and analysis
  methods,'' \emph{IEEE Journal of Selected Topics in Signal Processing},
  vol.~11, no.~1, pp. 48--63, Feb 2017.

\bibitem{Improve_analysis_2018}
J.~{LI} and P.~L. {CALLET}, ``Improving the discriminability of standard
  subjective quality assessment methods: a case study,'' in \emph{2018 Tenth
  International Conference on Quality of Multimedia Experience (QoMEX)}, May
  2018, pp. 1--3.

\bibitem{Freitas_2018_Performance}
\BIBentryALTinterwordspacing
P.~G. Freitas, A.~F. Silva, J.~A. Redi, and M.~C.~Q. Farias, ``Performance
  analysis of a video quality ruler methodology for subjective quality
  assessment,'' \emph{Journal of Electronic Imaging}, vol.~27, no.~5, pp. 1 --
  10 -- 10, 2018. [Online]. Available:
  \url{https://doi.org/10.1117/1.JEI.27.5.053020}
\BIBentrySTDinterwordspacing

\bibitem{User_Model_for_JND_2018}
\BIBentryALTinterwordspacing
H.~Wang, I.~Katsavounidis, X.~Zhang, C.~Yang, and C.-C.~J. Kuo, ``A user model
  for jnd-based video quality assessment: theory and applications,'' 2018.
  [Online]. Available: \url{https://doi.org/10.1117/12.2320813}
\BIBentrySTDinterwordspacing

\bibitem{Tasaka_2017_Bayesian_Hierarchical}
S.~{Tasaka}, ``Bayesian hierarchical regression models for qoe estimation and
  prediction in audiovisual communications,'' \emph{IEEE Transactions on
  Multimedia}, vol.~19, no.~6, pp. 1195--1208, June 2017.

\bibitem{Pablo_AMP_2019}
K.~D. {Singh}, Y.~{Hadjadj-Aoul}, and G.~{Rubino}, ``Quality of experience
  estimation for adaptive http/tcp video streaming using h.264/avc,'' in
  \emph{2012 IEEE Consumer Communications and Networking Conference (CCNC)},
  Jan 2012, pp. 127--131.

\bibitem{Romano_Correcting_2008}
R.~Romano, P.~B. Brockhoff, M.~Hersleth, O.~Tomic, and T.~Næs, ``Correcting
  for different use of the scale and the need for further analysis of
  individual differences in sensory analysis,'' \emph{Food Quality and
  Preference}, vol.~19, no.~2, pp. 197 -- 209, 2008, 8th Sensometrics Meeting.

\bibitem{Tomic_Visualization_2007}
O.~Tomic, A.~Nilsen, M.~Martens, and T.~Næs, ``Visualization of sensory
  profiling data for performance monitoring,'' \emph{LWT - Food Science and
  Technology}, vol.~40, no.~2, pp. 262 -- 269, 2007.

\bibitem{Maniscalco_SDT_2012}
B.~Maniscalco and H.~Lau, ``A signal detection theoretic approach for
  estimating metacognitive sensitivity from confidence ratings,''
  \emph{Consciousness and Cognition}, vol.~21, pp. 422--430, 2012.

\bibitem{Fleming_Self-evaluation_2017}
S.~Fleming and N.~D.~Daw, ``Self-evaluation of decision-making: A general
  bayesian framework for metacognitive computation,'' \emph{Psychological
  Review}, vol. 124, pp. 91--114, 01 2017.

\bibitem{King_A_model_of_subjective_2014}
J.-R. King and S.~Dehaene, ``A model of subjective report and objective
  discrimination as categorical decisions in a vast representational space,''
  \emph{Philosophical Transactions of The Royal Society B Biological Sciences},
  vol. 369, p. 20130204, 03 2014.

\bibitem{NUMBER_OF_RESPONSE}
D.~F. {Alwin}, E.~M. {Baumgartner}, and B.~A. {Beattie}, ``Number of response
  categories and reliability in attitude measurement,'' \emph{Journal of Survey
  Statistics and Methodology}, vol.~6, pp. 212--239, 2018.

\bibitem{PSYCHOMETRIC_TESTING}
D.~L. {Malott}, B.~R. {Fulton}, D.~{Rigamonti}, and S.~{Myers}, ``Psychometric
  testing of a measure of patient experience in {Saudi Arabia} and the {United
  Arab Emirates},'' \emph{Journal of Survey Statistics and Methodology},
  vol.~5, pp. 398--408, 2017.

\bibitem{Janowski2019NotationFS}
L.~Janowski, J.~Nawała, W.~Robitza, Z.~Li, L.~Krasula, and K.~Rusek,
  ``Notation for subject answer analysis,'' 2019.

\bibitem{betaBinomial}
\BIBentryALTinterwordspacing
K.~R. Coombes, \emph{The Beta-Binomial Distribution}. [Online]. Available:
  \url{https://cran.r-project.org/web/packages/TailRank/vignettes/betabinomial.pdf}
\BIBentrySTDinterwordspacing

\bibitem{HDTV_Phase_I_test}
M.~Pinson, F.~Speranza, M.~Barkowski, V.~Baroncini, R.~Bitto, S.~Borer,
  Y.~Dhondt, R.~Green, L.~Janowski, T.~Kawano \emph{et~al.}, ``Report on the
  validation of video quality models for high definition video content,''
  \emph{Video Quality Experts Group}, 2010.

\bibitem{ITS4S_test}
M.~H. Pinson, ``Its4s: A video quality dataset with four-second unrepeated
  scenes,'' NTIA/ITS, Tech. Rep. NTIA Technical Memo TM-18-532, Feb. 2018.

\bibitem{Janowski2014}
L.~{Janowski} and M.~{Pinson}, ``Subject bias: Introducing a theoretical user
  model,'' in \emph{2014 Sixth International Workshop on Quality of Multimedia
  Experience (QoMEX)}, Sep. 2014, pp. 251--256.

\bibitem{AGH_NTIA_14-505}
M.~H. Pinson and L.~Janowski, ``Agh/ntia: A video quality subjective test with
  repeated sequences,'' NTIA/ITS, Tech. Rep. NTIA Technical Memo TM-14-505,
  June 2014.

\bibitem{MM2_test}
M.~H. {Pinson}, L.~{Janowski}, R.~{Pepion}, Q.~{Huynh-Thu}, C.~{Schmidmer},
  P.~{Corriveau}, A.~{Younkin}, P.~{Le Callet}, M.~{Barkowsky}, and
  W.~{Ingram}, ``The influence of subjects and environment on audiovisual
  subjective tests: An international study,'' \emph{IEEE Journal of Selected
  Topics in Signal Processing}, vol.~6, no.~6, pp. 640--651, Oct 2012.

\bibitem{Walpole_Probablity_2007}
R.~E. Walpole, R.~H. Myers, S.~L. Myers, and K.~Ye, \emph{Probability \&
  statistics for engineers and scientists}, 8th~ed.\hskip 1em plus 0.5em minus
  0.4em\relax Upper Saddle River: Pearson Education, 2007.

\end{thebibliography}
\appendices

\section{Estimation}\label{sec:ap:est}

Numerically optimized likelihood function

\begin{equation*}
\ell(\mathbf u|\theta) =  \sum_i \begin{cases}
\ell_{\mathcal{B}}(u_{i}|\theta), \quad\rho\geq C(\psi)\\
\ell_{Bin}(u_{i}|\theta) + f(u_{i},\theta), \quad\rho<C(\psi),
\end{cases}
\end{equation*}
where
\begin{multline*}
	f(u_{i},\theta) =\log(1-\rho) - \log(1-C(\psi)) + \\ \log(1+  \frac{(\rho-C(\psi))[1-|u_{i}-\psi|]_{+}}{\binom{4}{u_{i}-1}\left(\frac{\psi-1}{4}\right)^{u_{i}-1}\left(\frac{5-\psi}{4}\right)^{5-u_{i}}(1-\rho)}),  
\end{multline*}
$\ell_{\mathcal{B}}(u_{i}|\theta)$ is a logarithm of $P_{G_{\rho}}(\epsilon=u_i-\psi)$ given by \eqref{eq:Prho} and $\ell_{Bin}$ is the likelihood function for Binomial distribution
 \begin{multline*}
 	\ell_{Bin}(u,\theta)= \log\binom{4}{u-1} + (u-1)(\log(\psi-1) -\log(4) ) +\\ (5-u)(\log(5-\psi) -\log(4)).
 \end{multline*}
Notice that the function $f$ is the only place where we do the computations in the probability domain (instead of log probability), however it is build around $\log(1+x)$ having numerically stable implementation in function \emph{log1p}. 

The logarithm of the newton symbol presents in both $\ell_{\mathcal{B}}$ and $\ell_{Bin}$ can also be efficiently implemented as 
\begin{multline*}
\log\binom{n}{k} = \log\Gamma(n+1) - \log\Gamma(k+1) - \log\Gamma(n-k+1),
\end{multline*}
where $\log\Gamma$ is a logarithm of the gamma function, provided as a special function \emph{lgamma} in TensorFlow. Notice that this component is commonly omitted from the likelihood because it does not depend on $\theta$. 

\section{Generalization}\label{sec:ap:n}

The described distribution is aiming $\{1, 2, \cdots, 5\}$ scale/support. Here we present a generalization for $\{1, 2, \cdots, M\}$ scale.

Let us denote 
$$V_{\mathrm{min}}(\psi)=(\lceil\psi\rceil-\psi)(\psi-\lfloor\psi\rfloor),$$
$$V_{\mathrm{max}}(\psi)=(\psi-1)(M-\psi),$$
$$C(\psi)=\frac{M-2}{M-1}\ \frac{V_{\mathrm{max}}(\psi)}{V_{\mathrm{max}}(\psi)-V_{\mathrm{min}}(\psi)}.$$
Let
\begin{equation} 
	\begin{split}
		P_{F_\rho} & (\epsilon=k-\psi)= \\ 
		& \frac{\rho-C(\psi)}{1-C(\psi)}[1-|k-\psi|]_{+} + \frac{1-\rho}{1-C(\psi)} \\
		& \binom{M-1}{k-1}\left(\frac{\psi-1}{M-1}\right)^{k-1}\left(\frac{M-\psi}{M-1}\right)^{M-k},
	\end{split}
\end{equation}
where $\rho\in[C(\psi),1]$, $k=1,...,M$ and
\begin{equation}
	\begin{split}
	P_{G_{\rho}}& (\epsilon=k-\psi)	= \binom{M-1}{k-1}\\
	& \frac{\mathcal{B}\left(\frac{(\psi-1)\rho}{(M-1)(C(\psi)-\rho)}+k-1,\frac{(M-\psi)\rho}{(M-1)(C(\psi)-\rho)}+M-k\right)}{\mathcal{B}\left(\frac{(\psi-1)\rho}{(M-1)(C(\psi)-\rho)},\frac{(M-\psi)\rho}{(M-1)(C(\psi)-\rho)}\right)}, 
	\end{split}
\end{equation}
where $\rho\in(0,C(\psi))$, $k=1,...,M$.

If we denote by $H_{\rho}$ the distribution function of the noise then we assume that
$$H_{\rho}=G_\rho\ I(\rho<C(\psi)) + F_\rho\ I(\rho\geq C(\psi)),$$
where $\rho\in(0,1]$ is a confidence parameter (see Remark \ref{rm}).
\begin{re}\label{rm}
	The variance of the noise is equal to
	$$
	\mathbb{V}_{H_{\rho}}(\epsilon)=\rho V_{\mathrm{min}}(\psi)+(1-\rho)V_{\mathrm{max}}(\psi).
	$$ 
	Since the variance of the noise is a decreasing function of $\rho\in(0,1]$ then this parameter has an interpretation as a confidence parameter. 
\end{re}

From this moment we assume the following
\begin{as}\label{eq:modM} 
	$$U=\psi+\epsilon,$$ 
	where $\psi\in[1,M]$ is an unknown parameter, $\epsilon$ are independent random variables with distribution functions $H_{\rho}$, where $\rho\in(0,1]$ is an unknown parameter.
\end{as}

\end{document}